\documentclass[twocolumn]{aastex62}
\pdfoutput=1 
\usepackage{amsmath,amstext}
\usepackage[T1]{fontenc}
\usepackage[figure,figure*]{hypcap}

\graphicspath{{./}{figures/}}
\usepackage{soul}
\sethlcolor{yellow}

\shorttitle{55 Cnc e}
\shortauthors{Nguyen et al.}

\begin{document}

\title{The Effect of Tidal Heating and Volatile Budgets on the Outgassed Atmosphere of 55 Cancri e}

\author[0009-0000-5587-7297]{Barron K. Nguyen}
\affiliation{Department of Earth and Planetary Sciences, Stanford University, Stanford, CA 94305-2115, USA}

\author[0000-0003-2915-5025]{Laura K. Schaefer}
\affiliation{Department of Earth and Planetary Sciences, Stanford University, Stanford, CA 94305-2115, USA}

\author[0000-0002-8958-0683]{Fei Dai}
\affiliation{Institute for Astronomy, University of Hawai`i, 2680 Woodlawn Drive, Honolulu, HI 96822, USA}

\author[0000-0002-8928-3929]{Héctor E. Delgado-Díaz}
\affiliation{Department of Astronomy, University of Washington, Seattle, WA 98195, USA}

\begin{abstract}\noindent 55 Cancri e is a $\sim$8 Gyr rocky world (1.95 $R_\oplus$, 8.8 $M_\oplus$) orbiting a K-type star. JWST observations suggest a carbon-dominated atmosphere (CO$_2$/CO) over a global magma ocean ($>$3000 K). We suggest that any CO$_2$-dominated atmosphere, with trace H$_2$O/O$_2$, likely arises from outgassing of its initial volatile reservoir. As solidification drives the magma ocean and atmosphere away from solution-equilibrium, tidal and greenhouse heating can prolong outgassing. Early atmosphere outgassing reflects rapid degassing of the volatile-saturated melt during post-formation cooling. Without tidal heating, an initial 5 wt\% water mass fraction ($F_{\text{H}_2\text{O}}$) or 3 wt\% $\text{CO}_2$ mass fraction ($F_{\text{CO}_2}$) can sustain outgassing for at least $\sim$10 Myr. With both at 10 wt\%, greenhouse warming alone can prolong outgassing up to $\sim$30 Myr. Our model shows that tidal heating can reduce the volatile threshold required to maintain a high surface temperature ($\sim$3200 K at $e = 0.005$) and delay outgassing of additional volatiles to the present-day. However, higher tidal heating presents a tradeoff between prolonging tenuous outgassing and enlarging the overall size of the secondary atmosphere. Tidally-enhanced outgassing may produce minor pressure variations that could contribute to the observed phase-curve variability. Additionally, our model shows that tidal heating strongly controls outgassing in the planet's young-to-midlife stage, then shifts toward a volatile inventory dependence at mature ages. Using 55 Cnc e, we present a framework to prioritize atmosphere detections on rocky ultra short period (USP) magma ocean planets, linking age-dependent tidal heating and volatile inventory to the formation and size of secondary atmospheres.
\end{abstract}



\section{Introduction}

55 Cancri e (55 Cnc e) is the fourth planet in the 55 Cnc A system \citep{butler_three_1997, mcarthur_detection_2004}, classified as an ultra-short period planet (USP) orbiting at or less than one day. Due to advances in detection, the number of USPs discovered has expanded significantly, comprising approximately $\sim$120 planets < 2$R_{\oplus}$ and < 10$M_{\oplus}$ broadly classified as terrestrial \citep{sanchis-ojeda_study_2014, dai_tks_2021}. Global surface temperatures can exceed the melting temperature of peridotite $\sim$1400 K at close orbits ($P_{\text{orb}}$ $\sim$0.74 days for 55 Cnc e), sufficient to form a global magma ocean. 

\begin{figure*}\label{fig:Mass_Radius_Magma_Ocean}
\center 
\includegraphics[width = 2.15\columnwidth]{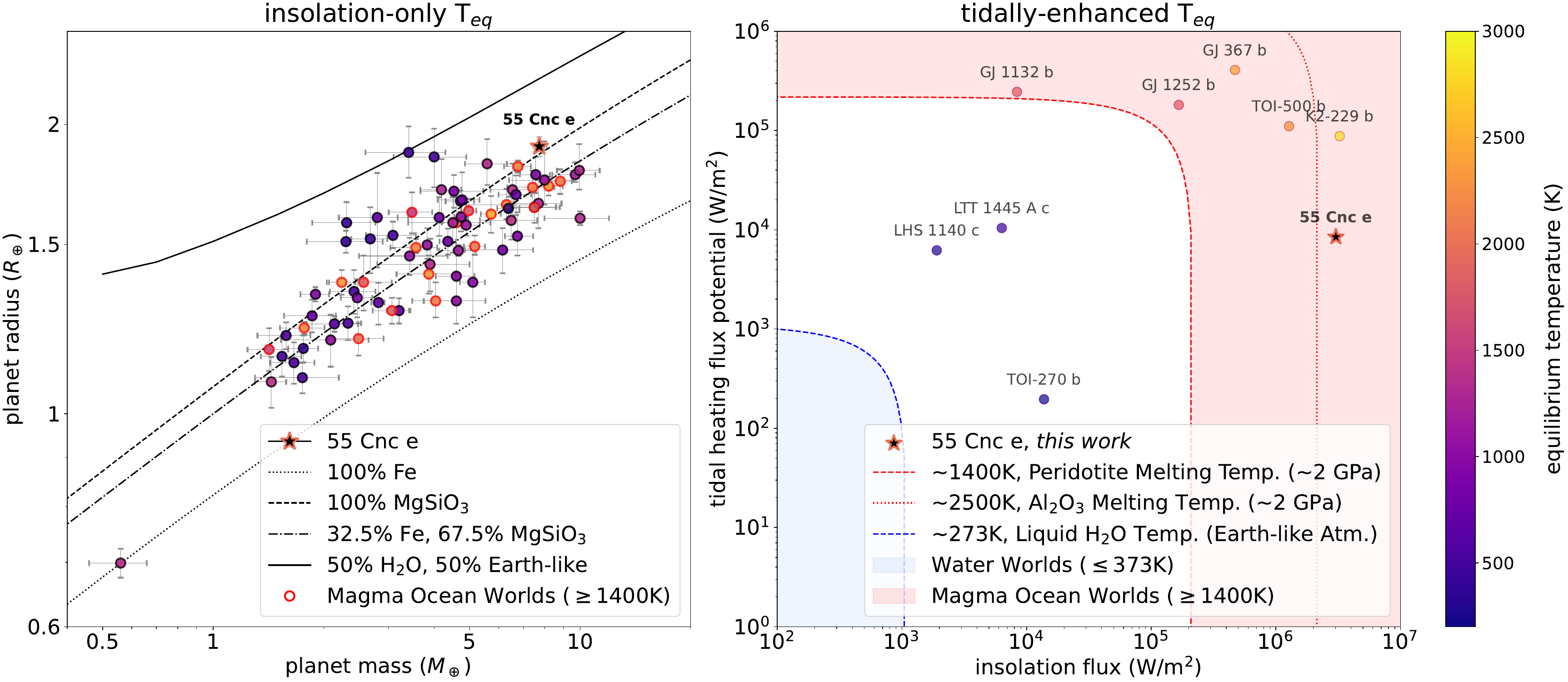}
\caption{ {\bf Left:} Mass-radius diagram of 75 ultra-short period planets (USPs) below 10 $M_{\oplus}$ and 2 $R_{\oplus}$ with both measured and constrained mass and radius, including 55 Cnc e which is marked with a star symbol. A USP planet is defined as a planet with an orbital period $P_{\text{orb}}$ < 1 day. The planetary mass and radii data are taken from the NASA Exoplanet Archive. Bulk planetary density curves from \citet{zeng_growth_2019} are overlaid, showing 100\% iron composition (dotted black), 100\% rocky composition (dashed black), Earth-like rocky composition (dot-dashed black), 50\% water vapor atmosphere with 50\% rocky composition (solid black), 5\% $\text{H}_2$ atmosphere with 95\% rocky composition at 1000 K (dotted gray), and 5\% $\text{H}_2$ atmosphere with 95\% rocky composition at 2000 K (dashed gray). The insolation-only equilibrium temperature $T_{\text{eq}}$ of each planet in K is plotted on a color scale, which is on the same scale as the tidally-enhanced equilibrium temperature $T_{\text{eq}}$ on the right. {\bf Right:} Tidal-insolation heat curve plot for planets included in our 75 USP dataset in which respective nominal tidal heating flux potentials were also calculated by \citet{mcintyre_s_r_n_tidally_2022}. The predicted effective $T_{\text{eq}}$ of planets with a tidal heating contribution,  from Eq. (\ref{eqn:tidal_Teq}) as $T_{\text{eq}}^{\text{tidal}}$, are shown on the same scale as the insolation-only equilibrium temperature. From this work, we evaluate the nominal value of tidal heating for 55 Cnc e as 8400 $\text{W m}^{-2}$, assuming the nominal eccentricity $e = 0.005$, and no atmosphere-derived greenhouse warming effect ($P_{\text{atm}} =0)$. Red and blue parameter spaces denote the effective equilibrium temperatures $T_{\text{eq}}$ that represent magma ocean worlds and water worlds respectively.}
\end{figure*}

Recent studies propose that tidal heating controls cooling timescales and can sustain observed volcanic atmospheres \citep{bolmont_emeline_tidal_2013,hammond_linking_2017, bello-arufe_evidence_2025, farhat_tides_2025, nicholls_self-limited_2025}. Tidal heating may be important in enhancing volatile outgassing \citep{jackson_tidal_2008} and influencing climate regulation \citep{reinhold_ignan_2025}. Tidal heating is therefore crucial for understanding the coupled thermal, volatile, and atmospheric evolution. Figure \ref{fig:Mass_Radius_Magma_Ocean} plots select rocky USPs on density and temperature curves, highlighting how tidal heating raises planetary temperatures. Additionally, constraining volatile reservoirs allows for better characterization of secondary atmosphere formation, planetary interiors, and bulk compositions \citep{elkins-tanton_ranges_2008, schmandt_dehydration_2014, dorn_hidden_2021, unterborn_nominal_2023, luo_interior_2024}. Magma ocean interactions with atmospheres have thus been a promising case-study of atmosphere-interior exchange, allowing for mantle evolution to be linked with volatile and atmosphere evolution \citep{elkins-tanton_linked_2008, hamano_emergence_2013, lebrun_thermal_2013, schaefer_predictions_2016, kite_atmosphere-interior_2016,kite_exoplanet_2020,kite_atmosphere_2020}.

Currently, 55 Cnc e is a hot and popular exoplanet to observe, offering a unique opportunity to probe the extremes of exoplanet evolution theory. Phase-curve variability of 55 Cnc e has been observed with TESS \citep{meier_valdes_weak_2022}, CHEOPS \citep{morris_cheops_2021, demory_55_2023, meier_valdes_investigating_2023}, Spitzer's IRAC \citep{demory_variability_2016, tamburo_confirming_2018, mercier_revisiting_2022}, and JWST's NIRCam \citep{patel_jwst_2024}. Retrieved surface temperatures vary between dayside and nightside observations. Spitzer/IRAC studies estimate dayside–nightside temperatures ranging from ~1400–2700 K \citep{demory_map_2016} to ~1700–3800 K \citep{mercier_revisiting_2022}, while \citet{hu_secondary_2024} measured a brightness temperature of $\sim$1800 K using JWST, suspecting atmospheric heat redistribution. Models for atmospheric composition have shifted towards heavy molecular weight atmospheres with no H/He, based on non-detections of those species using Keck's NIRSPEC \citep{zhang_no_2021}, GTC's HORuS \citep{tabernero_horus_2020}, and most recently with JWST's NIRCam/MIRI \citep{hu_secondary_2024, zilinskas_characterising_2025}, which are improvements from previous atmosphere non-detections using CARMENES \citep{deibert_near-infrared_2021} and Gemini North's MAROON-X \citep{rasmussen_nondetection_2023}. A rocky-silicate vapor atmosphere has also been ruled out with LBT's PEPSI \citep{keles_pepsi_2022} and JWST \citep{hu_secondary_2024}, though previously inferred by VLT's UVES and HARPS \citep{ridden-harper_search_2016}. At present, observations made with JWST's NIRCam/MIRI \citep{hu_secondary_2024, zilinskas_characterising_2025} suggest a CO and or $\text{CO}_2$-dominant atmosphere. Complementary observations by Subaru's HDS/CFHT \citep{esteves_search_2017}, Gemini North \citep{jindal_characterization_2020}, and Hubble \citep{tsiaras_detection_2016}, have hinted at possible trace background gases like $\text{H}_2\text{O}$ and $\text{HCN}$, among others. Eccentricity estimates range from $\sim$0.05, from combined RV data including SOPHIE \citep{bourrier_v_55_2018}, to $\sim$0.03 from Keck/Lick RV data \citep{nelson_55_2014}, and $\sim$0.01 from Spitzer secondary eclipse fits \citep{demory_detection_2012, bolmont_emeline_tidal_2013}. In contrast, model-derived estimates lower eccentricities ($e<0.01$) owing to short circularization timescales and extreme tidal heating \citep{bolmont_emeline_tidal_2013, baluev_enhanced_2015, demory_map_2016, hammond_linking_2017, mello_tidal_2025}. One exciting capability of JWST will be its ability to place stringent constraints on eccentricity through precise timing of future secondary eclipses \citep{huber_discovery_2017}.

Broadly speaking, the interior-atmosphere feedback, magma ocean behavior, and orbital dynamics of 55 Cnc e have been extensively modeled separately \citep{madhusudhan_possible_2012, bolmont_emeline_tidal_2013, hammond_linking_2017, crida_mass_2018, kite_atmosphere_2020, boukare_deep_2022, heng_transient_2023, piette_rocky_2023, farhat_tides_2025, loftus_extreme_2025, mello_tidal_2025, zilinskas_characterising_2025}. In this paper, we investigate the thermal evolution and outgassing history of 55 Cnc e by applying a magma ocean evolution model modified from \citet{schaefer_predictions_2016} to include carbon dioxide and tidal heating effects. Our results provide new insights into the outgassing history and the driving factors for 55 Cnc e and similar extreme close-in magma ocean worlds. In Section \ref{sec:atm_model}, we first discuss the stellar and atmosphere models, followed by the interior model in Section \ref{sec:int_model}. In Section \ref{sec:results}, we discuss the results of our models, both those with and without tidal heating, and in Section \ref{sec:discussion}, we discuss some of our model limitations and implications of our results for 55 Cnc e and USPs more generally. 

\section{Atmosphere and Climate Model}\label{sec:atm_model}

In the following section, we present our implementation of an atmosphere and climate model based on the model of \citet{schaefer_predictions_2016}. Section \ref{subsec:stellar_climate_model} outlines the adopted stellar parameterizations for a Sun-like star, which drives the predicted atmospheric dissociation and loss on 55 Cnc e, as well as its high equilibrium temperature. In Section \ref{subsec:hydrodynamic_loss_model}, we extend the XUV-driven hydrodynamic loss model of \citet{schaefer_predictions_2016} to include $\text{CO}_2$ in order to assess how high molecular weight atmospheres may evolve due to the extreme atmospheric loss environments of an ultra close-in terrestrial planet. We describe the planet interior model in Section \ref{sec:int_model}.

\subsection{Stellar and Climate Model}\label{subsec:stellar_climate_model}

55 Cnc e orbits 55 Cnc A, which is a K-type star with $T_{\text{eff}}=5272 \pm 24$ K \citep{zhao_measured_2023}, and which is the primary star in the 55 Cnc binary system. We employ a simple climate model that calculates the outgoing longwave radiation (OLR) of 55 Cnc e's predicted atmosphere, based on its atmospheric properties and surface temperature evolution. The insolation-only equilibrium temperature $T_{\text{eq}}$ is given by the equation:

\begin{deluxetable}{lccc} 
\tablecaption{Stellar and Model Parameters of 55 Cnc A (HR-3522) 
\label{tab:stellar_para}}
\tablewidth{0.8\columnwidth} 
\tablehead{
\colhead{Parameter} & \colhead{Value} & \colhead{Unit} & \colhead{Reference}
}
\startdata
TIC ID & 332064670 & & 1 \\
Spec. T & K0IV-V / G8V & & 2, 3 \\ 
$T_{\text{eff}}$ & 5250 & K & 1 \\
$g$ & 4.4245900 & $\log~g$ (cm s$^{-2}$) & 1 \\
$M_{\star}$ & $	0.905\pm0.015$ & $M_{\odot}$ & 4 \\
$R_{\star}$ & $0.943\pm0.010$ & $R_{\odot}$ & 4 \\
$L_\star$ & $0.640$ & $L_{\odot}$ & 1, 5 \\
$f_0$ & 0.001 & & 6 \\
$\beta_\text{XUV}$ & --1.23 & & 7, 8, 9 \\
$t_{\text{sat}}$ & 1e8 & yr & 9 \\
$\epsilon_{\text{XUV}}$ & 0.3 &  & 10, 11 \\
\enddata

\tablecomments{1. TICv8; 2. \citet{braun_55_2011}; 3. \citet{butler_three_1997}; 4. \citet{bourrier_v_55_2018}; 5. \citet{gaia_collaboration_gaia_2018}; 6. \cite{lehmer_rocky_2017}; 7. \citet{luger_extreme_2015}; 8. \citet{hu_secondary_2024}; 9. \citet{ribas_evolution_2005}; 10. \citet{tian_thermal_2009}; 11. \citet{schaefer_predictions_2016}.}
\end{deluxetable}

\begin{equation}\label{eqn:Teq}
T_{\text{eq}} = \left( \frac{(1 - A) L_\star}{16 \pi \sigma a^2} \right)^{\frac{1}{4}}
\end{equation}

\noindent where $A=0.1$ is the bond albedo. We adopt the nominal value suggested by \citet{essack_low-albedo_2020} to be the upper estimate of a planetary-wide solid quenched glass surface or global liquid magma ocean. $L_{\star}$ is the stellar luminosity for a 0.9 M$_{\odot}$ star from \citet{baraffe_new_2015}, and $a$ is the semi-major axis, which is calculated from the measured orbital period using Kepler's third law. Nominal stellar and planetary parameter values used to calculate the insolation-only equilibrium temperature, which we distinguish from the isolation$+$tidal heating value, in Eq. (\ref{eqn:Teq}) can be found in Table \ref{tab:stellar_para} and Table \ref{tab:planet_para} respectively. 

The net outgoing longwave radiation (OLR) flux is dependent on the evolution of the insolation-only equilibrium temperature as well as the surface temperature $T_{\text{surf}}$, given by

\begin{equation}\label{eqn:OLR_flux}
\mathcal{F}_{\text{OLR}} = \varepsilon \sigma \left( T^4_{\text{surf}} - T^4_{\text{eq}} \right)
\end{equation}

\noindent where $T_{\text{surf}}$ is the surface temperature governed by the internal thermal evolution model outlined in Section \ref{sec:int_model}, $\sigma$ is the Stefan–Boltzmann constant, and $\varepsilon$ is the emissivity. We use a gray emitter approximation for the atmosphere as parameterized from \citet{elkins-tanton_linked_2008},

\begin{equation}\label{eqn:emissivity}
\varepsilon = \frac{2}{\tau +2}
\end{equation}

\noindent where $\tau$ is the whole atmosphere optical depth calculated as a function of the abundance of each species $i$ \citep{abe_formation_1985},

\begin{equation}\label{eqn:tau_optical_depth}
\tau = \sum \frac{3 M_{\text{atm},i}}{8 \pi R_p^2} \sqrt{\frac{k_{0,i} g}{3 p_0}}
\end{equation}

\noindent where $M_{\text{atm},i}$ is the mass of species $i$ in the atmosphere, $R_p$ is the planetary radius, $g$ is the gravitational acceleration, and $k_{0,i}$ is the absorption coefficient of species $i$ at a standard reference pressure $p_0$. From previous work, we adopt $k_{0,\text{CO}_2} = 0.001 \text{ m}^2\text{ kg}^{-1}$ and $k_{0,\text{H}_2\text{O}} = 0.01 \text{ m}^2\text{ kg}^{-1}$ \citep{yamamoto_radiation_1952,abe_evolution_1988,nikolaou_what_2019}. OLR calculations are used to determine the thermal equilibrium between the interior heat flux and surface heat loss. We include the effects of tidal heating in Section \ref{sec:int_model}.

\subsection{XUV-Driven Hydrodynamic Atmospheric Loss}\label{subsec:hydrodynamic_loss_model}

In our atmospheric loss model, we assume XUV-driven hydrodynamic loss to be the main mechanism contributing to the escape of CO$_2$, H$_2$O, and O$_2$ in the atmosphere. The atmospheric mass loss rate is a function of the stellar flux received at 55 Cnc e's orbital distance, and is primarily assumed to be driven by hydrogen produced by the dissociation of outgassed $\text{H}_2\text{O}$. Heavier molecular species such as $\text{CO}_2$, alongside atomic oxygen produced from dissociated $\text{H}_2\text{O}$, are expected to be dragged away by escaping hydrogen at varying efficiencies, potentially limiting the formation of a substantial secondary atmosphere unless replenished by continuous outgassing \citep{kite_exoplanet_2020, krissansen-totton_erosion_2024}. We explore how tidal heating effects outgassing rates from the magma ocean in Section \ref{subsec:tidal_model} and \ref{sec:results_tidal_outgassing}, where we show that it leads to continuous albeit tenuous outgassing on Myr-Gyr timescales.

For Sun-like stars, the XUV saturation fraction $f_0$ of stellar bolometric luminosity $L_{\text{bol}}$ is generally assumed to be $\sim 0.1\%$, representing a high XUV flux that can drive hydrodynamic escape of an atmosphere \citep{ribas_evolution_2005}. We adopt the bolometric luminosity ($L_{\text{bol}}$) evolution model from \citet{baraffe_new_2015} and the model of \citet{ribas_evolution_2005} for the fractional XUV luminosity of 55 Cnc A, described in the form of a power law as a function of time,

\begin{equation}\label{eqn:XUV_luminosity}
L_{\text{XUV}} =
\begin{cases}
L_{\text{bol}} f_0 &  \text{, } t < t_{\text{sat}} \\
L_{\text{bol}} f_0 \left( \dfrac{t}{t_{\text{sat}}} \right)^{\beta_{\text{XUV}}} & \text{, } t \geq t_{\text{sat}}
\end{cases}
\end{equation}

\noindent where $\beta_\text{XUV}$ is the decay constant, $t$ is the time after star formation (yr), $f_0$ is the XUV saturation fraction, and $t_{\text{sat}}$ is the saturation time (yr). The XUV fraction is held constant at $f_0$ until the saturation time, and thereafter undergoes decay. We use $t_{\text{sat}}=100\text{ Myr}$, corresponding to the duration of heightened XUV emission from young Sun-like stars \citep{ribas_sun_2009}. A range-averaged value, from X-ray to UV, is implemented to produce the combined luminosity decay constant of $\beta_\text{XUV}=-1.23$ \citep{ribas_evolution_2005}, though separate time evolution models for X-ray ($\beta_\text{X-ray}=-0.9$) versus UV ($\beta_\text{XUV}=-1.3$) wavelengths have also been used to model escape rates based on the extreme ultraviolet flux for 55 Cnc e \citep{hu_secondary_2024}.

Incorporating the expected stellar XUV contribution over time, the resulting overall mass escape rate ($\text{kg}$ $\text{m}^2$ $\text{s}^{-1}$) for all volatile species in the atmosphere ($\phi$) can be described by the following equation, as adopted from \citet{zahnle_mass_1990},

\begin{equation}\label{eqn:overall_mass_escape}
\phi = \frac{\epsilon_{\text{XUV}} F_{\text{XUV}}}{4V_{\text{pot}}}
\end{equation}

\noindent in which $\epsilon_{\text{XUV}}$ represents the XUV energy conversion efficiency factor that considers atmospheric absorption efficiencies and its altitude-dependent variability, and $F_{\text{XUV}}$ is the XUV radiation flux at the orbit of 55 Cnc e. We take the nominal value of $\epsilon_{\text{XUV}}$ to be the upper estimate of 0.3 per the conservative stellar flux model we employ, which results in a higher mass loss rate of the secondary atmosphere \citep{tian_thermal_2009, schaefer_predictions_2016}. $V_{\text{pot}}$ is the gravitational potential ($V_{\text{pot}} = G M_{p}/R_{p}$) using the planetary parameters for 55 Cnc e in Table \ref{tab:planet_para}.

We adopt an energy-limited escape formulation for atmospheric escape, where we assume that atomic H, derived from photolysis of water vapor, drives the outflow. We use the cross-over mass model of \cite{hunten1987mass} to determine the rate of escape of heavier species that are entrained in the hydrogen outflow. We do not explicitly calculate the photolysis rate of water vapor, but assume complete dissociation to hydrogen and oxygen atoms. We also do not consider the photolysis of $\text{CO}_2$ to CO and atomic oxygen due to possible efficient recombination via catalytic cycles \citep{ranjan_photochemistry_2020}. Past studies have shown the photolysis rate of $\text{CO}_2$ to be three orders of magnitude lower than $\text{H}_2\text{O}$ dissociation into H and OH on habitable rocky exoplanets (295 K) orbiting G stars, due to the stronger double bond strength of C and O in molecular $\text{CO}_2$ \citep{hu_photochemistry_2012, harman_abiotic_2018}.

The rate of atmospheric escape of heavy species such as $\text{CO}_2$ and $\text{O}_2$ depends on their mass and the hydrogen escape flux. We calculate the crossover mass, which is the mass of the heaviest species that can escape. The  reference crossover mass is:

\begin{equation}\label{eqn:ref_crossover_mass}
\mu_{c_i}^{\text{ref}} = \mu_\text{H} + \frac{k_B T_{\text{esc}} \Phi_{\text{H}}^{\text{ref}}}{b_{\text{H},i} g X_\text{H} m_{\text{p}}}
\end{equation}

\noindent and the actual crossover mass after calibration, as adopted from \citet{chassefiere_hydrodynamic_1996} is,

\begin{equation}\label{eqn:crossover_mass_correction}
\mu_{c_{i}} = \mu_i + \gamma_{\text{H},i} (\mu_{c_i}^{\text{ref}} - \mu_i)
\end{equation}

\noindent where $\mu_{c_i}^{\text{ref}}$ is the reference threshold crossover mass for secondary species $i$, $\mu_i$ and $\mu_\text{H}$ are the molecular masses of species $i$ and hydrogen respectively, $k_B$ is the Boltzmann constant, $T_{\text{esc}}$ is the temperature of the escaping region, $\Phi_{\text{H}}^{\text{ref}}$ is the reference hydrogen escape rate (see Eq. (\ref{eqn:ref_hydrogen_flux})), $b_{\text{H},i}$ is the binary diffusion coefficient for hydrodynamic drag of the heavier species $i$ (from \citet{zahnle_mass_1986}, $b_{\text{H},\text{O}} = 4.8 \times 10^{19} T_{\text{esc}}^{0.75}$ $\text{m}^{-1} \text{ s}^{-1}$, and $b_{\text{H},\text{CO}_2} = 8.4 \times 10^{19}T_{\text{esc}}^{0.6}$ $\text{m}^{-1} \text{ s}^{-1}$), $X_\text{H}$ is the molar concentration of hydrogen, and $m_p$ is the proton mass. For Eq. (\ref{eqn:crossover_mass_correction}), $\gamma_{\text{H},i}$ is the mixing ratio calibration factor of species $i$ being dragged by escaping hydrogen which is calculated by:

\begin{equation}\label{eqn:gamma_calibration}
\gamma_{\text{H},i} = \frac{1}{1 + \frac{X_i \mu_i}{X_\text{H} \mu_\text{H}}}
\end{equation}

\noindent using the molar concentrations and molecular masses of the species \citep{chassefiere_hydrodynamic_1996}. The reference flux of hydrogen, $\Phi_{\text{H}}^{\text{ref}}$ ($\text{molecules}$ $\text{m}^{-2}$ $\text{sec}^{-1}$), is the escape rate of hydrogen in the absence of other species and is given by:

\begin{equation}\label{eqn:ref_hydrogen_flux}
\Phi_{\text{H}}^{\text{ref}} = \frac{\phi}{\mu_\text{H} m_p}
\end{equation}

\noindent where the generalized mass flux $\phi$ is taken from Eq. (\ref{eqn:overall_mass_escape}). In a mixed binary composition atmosphere, the flux of hydrogen is dependent on the abundances and resulting diffusion-limited loss rates of the heavier species,

\begin{equation}\label{eqn:real_hydrogen_flux}
\Phi_{\text{H},i} = \Phi_{\text{H}}^\text{ref} \left( \frac{\mu_{c_i}}{\mu_{c_i}^{\text{ref}}} \right)
\end{equation}

The minimum amount of hydrogen flux $\Phi_{\text{H},i}^{\text{crit}}$ needed to power the escape of heavier species is described by the equation:

\begin{equation}\label{eqn:crit_hydrogen_flux}
\Phi_{\text{H},i}^{\text{crit}} = \frac{b_{\text{H},i} g X_\text{H} m_p (\mu_i - \mu_\text{H})}{k_B T_{\text{esc}}}.
\end{equation}

\noindent Note that we calculate the hydrogen fluxes needed to surpass the critical hydrogen flux thresholds for atmospheric mass loss of $\text{CO}_2$ and atomic oxygen separately because of the binary nature of known diffusion coefficients \citep{zahnle_mass_1986}.

If adequate hydrogen flux is present, the resulting fluxes ($\text{molecules}$ $\text{m}^{-2}$ $\text{sec}^{-1}$) of hydrogen, $\text{CO}_2$ and atomic oxygen can be represented as,

\begin{equation}\label{eqn:mass_flux}
\begin{aligned}
\Phi_{\text{H},i} &=
\begin{cases} 
\Phi_{\text{H}}^\text{ref} \left( \frac{\mu_{c_i}}{\mu_{c_i}^{\text{ref}}} \right) & , P_{\text{H}_2\text{O}} \geq P_{i} \\[10pt]
\Phi_{\text{H},i}^{\text{crit}} & , P_{\text{H}_2\text{O}} < P_{i} 
\end{cases} \\[10pt]
\Phi_i &=
\begin{cases} 
\Phi_{\text{H},i} \dfrac{X_i}{X_\text{H}} 
\frac{\mu_{c_i} - \mu_i}{\mu_{c_i} - \mu_\text{H}} 
& ,\begin{array}{c} \Phi_{\text{H},i} \geq \Phi_{\text{H},i}^{\text{crit}} \\ P_{\text{H}_2\text{O}} > P_{i} \end{array} \\[10pt]
0 & ,\begin{array}{c} \Phi_{\text{H},i} < \Phi_{\text{H},i}^{\text{crit}} \\ \text{or } P_{\text{H}_2\text{O}} \leq P_{i} \end{array} 
\end{cases}
\end{aligned}
\end{equation}

\noindent depending on atmospheric abundances $P_{\text{H}_2\text{O}}$, $P_{\text{CO}_2}$ and $P_{\text{O}_2}$, and the hydrogen flux rate at a specific time, $\Phi_{\text{H},i}$. We assume hydrodynamic mass loss of the heavier species to only operate under a steam-dominated atmosphere where the dissociation of $\text{H}_2\text{O}$ is sufficient to provide a hydrogen flux exceeding the minimum critical hydrogen flux, $\Phi_{\text{H},i}^{\text{crit}}$ \citep{luger_extreme_2015, schaefer_predictions_2016}.

\section{Interior Evolution Model}\label{sec:int_model}
In this section, we discuss the internal structure model, beginning with the thermal evolution model. In Section \ref{subsec:initialvolatiles}, we discuss the initial volatile budget and the volatile evolution model. In Section \ref{subsec:tidal_model}, we describe the tidal heating model that we use and the eccentricity values that we adopt for 55 Cnc e. 

\subsection{Planetary Mantle and Thermal Model}

Our interior thermal evolution model is a modification of the model presented in \citet{schaefer_predictions_2016}. Mantle temperature is controlled by a balance of heat production and heat loss processes. Thermal inputs from tidal heating ($q_{\text{tidal}}$) and radiogenic heat ($q_{\text{rad}}$) are parameterized as heat production processes in the interior, with a convection model controlling the mantle heat loss to the surface. We discuss the parameterized tidal heating model further in Section \ref{subsec:tidal_model}. The radiogenic heating is based on elemental abundances in an Earth-like planet and is given by:

\begin{equation}\label{eqn:Q_radiogenic}
q_{\text{rad}} = M_{\text{mantle}} \left( \sum\ C_{j} H_{j} e^{\lambda_j(t_{\text{present}}-t)}  \right)
\end{equation}

\noindent where we adopt values of isotopes $j$, for ${}^{238}\mathrm{U}$, ${}^{235}\mathrm{U}$, ${}^{232}\mathrm{Th}$, and ${}^{40}\mathrm{K}$ due to their long-lived nature, and ignore short-lived radionuclides such as ${}^{26}\mathrm{Al}$ due to their negligible heat contribution to the long-term global magma ocean state \citep{chao_lava_2021}. For each isotope $j$, $C_{j}$ is the radioisotope abundance in parts per billion relative to total uranium, $H_{j} $ is the radioactive heat constant in W $\text{kg}^{-1}$ (from \citet{schubert_mantle_2001}, and $\lambda_j$ is the decay constant in $\text{yr}^{-1}$ (from \citep{mcnamara_cooling_2000} (see Table \ref{tab:planet_para}. The remaining parameters, $M_{\text{mantle}}$ and $(t_\text{present}-t)$, describe the mass of the mantle and the elapsed time since the formation of the planet, respectively. 

The planetary interior and mantle are assumed to be wholly convective and well-mixed. The convective zone depth $L_{\text{conv}}$ is given by:

\begin{equation}\label{eqn:convective_depth_L}
L_{\text{conv}} = \begin{cases}
R_p - r_s & \text{, } \phi_{\text{melt}} \geq 0.4 \\
R_p - R_c & \text{, } \phi_{\text{melt}} < 0.4 \\
\end{cases}
\end{equation}

\noindent where for bottom-up solidification of a magma ocean \citep{monteux_cooling_2016}, $R_p$ is the planetary radius, $r_s$ is the radius of solidification, taken as the radius of the solid portion of the mantle, and $R_c$ is the radius of the core ($R_c = f_{\text{CRF}} R_p$, see Table \ref{tab:planet_para}). During the magma ocean stage, we assume that convection occurs in the layer in which the melt fraction $\phi_{\text{melt}}$ is $>\phi_{\text{melt}}^{\text{crit}} (=0.4)$, the critical melt fraction, below which the viscosity transitions from a solid-like to liquid-like regime as described by \citet{lebrun_thermal_2013}. For scenarios where $\phi_{\text{melt}} < 0.4$, a substantial increase in mantle viscosity is expected to result in solid-like behavior, defining the start of solid-state convection. We assume that whole mantle convection occurs in this regime. 

We calculate the melt fraction as the mantle-averaged value over the partially molten region,

\begin{equation}\label{eqn:melt_frac_global}
{\phi_{\text{melt}}} = \frac{3}{R_{p}^3 - r_{s}^3} \int_{r_{s}}^{R_{p}} r^2\phi_{\text{melt}}(r)  \, dr
\end{equation}

\noindent using the boundaries of the magma ocean, $r_s$ and $R_p$. The local melt fraction $\phi_{\text{melt}}(r)$ is determined by the relation:

\begin{equation}\label{eqn:melt_frac_local}
\phi_{\text{melt}}(r) = \frac{T_{\text{mantle}}(r) - T_{\text{sol}}(r)}{T_{\text{liq}}(r) - T_{\text{sol}}(r)}
\end{equation}

\noindent in which $T_{\text{sol}}$ is the solidus temperature, $T_{\text{liq}}$ is the liquidus temperature, and $T_{\text{mantle}}$ is the adiabatic temperature profile of the mantle. The solidus and liquidus temperatures are taken from \citet{schaefer_predictions_2016}, which uses a linearization of the solidus from \citet{hirschmann_mantle_2000}, and are given by:

\begin{equation}\label{eqn:T_solidus}
T_{\text{sol}}(P) = 
\begin{cases}
104.42 P + 1420 & \text{, } P \leq 5.20\ \text{GPa} \\
26.53 P + 1825 & \text{, } P > 5.20\ \text{GPa}
\end{cases}
\end{equation}

\noindent as a function of pressure ($P$) in the planetary interior. We assume the mantle liquidus is a constant 600 K above the solidus temperature. 

The rate of mantle convection model is expressed by the Rayleigh number, which is given by:

\begin{equation}\label{eqn:Rayleigh_number}
\text{Ra} = \frac{g \, \alpha \, (T_{\text{mantle}} - T_{\text{surf}}) \, L_{\text{conv}}^3}{\nu \, \kappa}
\end{equation}

\noindent where $\alpha$ is the coefficient of thermal expansion, $\kappa$ is thermal diffusivity, and $\nu$ is the kinematic viscosity (see Table \ref{tab:planet_para}). 

The kinematic viscosity is given by the ratio of the dynamic viscosity (Pa s) to the mantle density. The reference dynamic viscosities ($\eta$) of the solid \citep{mcgovern_thermal_1989} and liquid \citep{karki_viscosity_2010,lebrun_thermal_2013} silicate can be written using the Arrhenius form, and are given by:

\begin{equation}\label{eqn:liq_dynamic_viscosity}
\eta_l =A  \exp{\left( \frac{B}{T_{\text{mantle}} - 1000} \right)}
\end{equation}

\begin{equation}\label{eqn:sol_dynamic_viscosity}
\eta_s = C  \exp{\left( \frac{D}{R T_{\text{mantle}}} \right)}
\end{equation}

\noindent where $T_{\text{mantle}}$ is the mantle temperature, and the empirically-derived constants are given in Table \ref{tab:planet_para}. The resulting kinematic viscosity then exhibits a dependence on the melt fraction, where the viscosity below the critical melt fraction $\phi^{crit}_{melt}$ has been shown to be liquid-like in contrast to the solid-like viscosity above the critical melt fraction \citep{lebrun_thermal_2013}:

\begin{equation}\label{eqn:kinematic_viscosity}
\nu =
\begin{cases}
\frac{\eta_l}{\rho_m} \left( 1 - \frac{1 - \phi_{\text{melt}}}{1 - \phi_{\text{melt}}^{\text{crit}}} \right)^{-2.5} & \text{, } \phi_{\text{melt}} \geq \phi_{\text{melt}}^{\text{crit}} \\
\frac{\eta_s}{\rho_m}  e^{\left( -\alpha_n \phi_{\text{melt}} \right)} & \text{, } \phi_{\text{melt}} < \phi_{\text{melt}}^{\text{crit}}
\end{cases}
\end{equation}

\noindent $\alpha_n$ is a diffusion creep constant from \citet{mei_influence_2002}. 

The heat flux $q_{\text{mantle}}$ from the mantle to the surface is then controlled by the recursive feedback between melt fraction, viscosity, and the corresponding convective regime:

\begin{equation}\label{eqn:Q_mantleflux}
q_{\text{mantle}} = \text{Nu}\frac{k_m(T_{\text{mantle}} - T_{\text{surf}})}{L_{\text{conv}}}
\end{equation}

\noindent where $\text{Nu}$ is the Nusselt number parameterized in the form:

\begin{equation}\label{eqn:Nusselt}
\text{Nu}=C_0(\text{Ra})^{\beta}
\end{equation}

\noindent where we adopt $C_0=0.089$ and $\beta=1/3$ \citep{lebrun_thermal_2013}. The thickness of a thin conductive thermal boundary layer at the top of the convective layer can then be expressed as:

\begin{equation}\label{eqn:thermal_boundary}
\delta_b = \frac{k_m (T_{\text{mantle}} - T_{\text{surf}})}{q_{\text{mantle}}}
\end{equation}

\noindent The thickness of the thermal boundary layer increases slowly within the fully liquid regime of the magma ocean, when it is millimeters to centimeters in size. When the melt fraction of the magma ocean reaches the critical melt fraction and viscosity jumps to solid-like values, the thermal boundary layer grows suddenly to up to kilometers in size. 

We begin simulations with a starting mantle temperature $T_{\text{mantle},i}$ of $\sim$4000 K, based on an approximate temperature $T_{\text{CMB},i}$ required for melting to occur up to the pressure of Earth's core-mantle boundary \citep{fiquet_melting_2010, andrault_solidus_2011}. The evolution of the mantle temperature $T_{\text{mantle}}$, radius of solidification $r_s$, and surface temperature $T_{\text{surf}}$ follows \citet{schaefer_predictions_2016}:

\begin{equation}\label{eqn:mantle_temp}
\frac{dT_\text{mantle}}{dt} = 
\frac{
    -4\pi R_p^2 q_{\text{mantle}} + q_{\text{rad}} + q_{\text{tidal}}
}{
    4\pi \rho_m c_{p,m} 
    \left[
        \frac{1}{3}(R_p^3 - r_s^3) 
        - \Delta H_f \left( r_s^2 \frac{dr_s}{dT_{\text{mantle}}} \right)
    \right]
}
\end{equation}

\begin{equation}\label{eqn:radius_of_solidification}
\frac{dr_s}{dt} = \frac{dT_{mantle}}{dt}\frac{c_{p,m} \left( b \alpha - a \rho_m c_{p,m} \right)}{g \left( a \rho_m c_{p,m} - \alpha T_{\text{mantle}} \right)^2}
\end{equation}

\begin{equation}\label{eqn:surf_temp}
\frac{dT_\text{surf}}{dt} =
4 \pi R_p^2
\left[
\frac{
q_{\text{mantle}}-\mathcal{F}_{\text{OLR}}
}{
\left( \sum c_{p,i} P_{\mathrm{i}}  \frac{4 \pi R_p^2}{g} \right) +
\frac{4}{3} \pi c_{p\text{,m}}  \rho_\text{m} \left(R_p^3 - \delta_b^3
\right)
}
\right]
\end{equation}

\noindent where $q_{tidal}$ is the tidal heat flux, described in Section \ref{subsec:tidal_model}, $\Delta H_f$ is the heat of fusion, or latent heat of melting, of silicate. The radius of solidification is directly coupled to the evolving mantle temperature from Eq. (\ref{eqn:mantle_temp}), where $a$ and $b$ are the linear fit coefficients of the solidus $T_{\text{sol}}(P)$ in Eq. (\ref{eqn:T_solidus}). The mantle temperature dictates the resulting surface temperature, which is dependent on the atmospheric composition, specifically the partial pressure $P_i$ and specific heat capacities $c_{p,i}$ of each species $i$, as well as the boundary layer thickness $\delta_b$ from Eq. (\ref{eqn:thermal_boundary}). These three equations, Eq. (\ref{eqn:mantle_temp}), (\ref{eqn:radius_of_solidification}), and (\ref{eqn:surf_temp}) describe the internal thermal evolution model we employ in modeling a global magma ocean.

\begin{deluxetable}{lccc} 
\tablecaption{55 Cnc e Planetary and Model Parameters \label{tab:planet_para}} 
\tablewidth{0.8\columnwidth} 
\tablehead{
\colhead{Parameter} & \colhead{Value} & \colhead{Unit} & \colhead{Reference}
}
\startdata
$M_{p}$ & 8.8 & $M_{\oplus}$ & 1, 2 \\
$R_{p}$ & 1.95 & $R_{\oplus}$ & 1, 2 \\
$P_{\text{orb}}$ & 0.7365474 & days & 10 \\
$a$ & 0.01582 & AU & 10 \\
$\tau_{\text{age}}$ & 7.4 - 8.7 & Gyr & 1, 3 \\
$P_{\text{CMB}}$  & 1200 & GPa & 4 \\
$f_{\text{CRF}}$ & 0.36 & constant & 2 \\
$f_{\text{CMF}}$ & 0.325 & constant & 13 \\
$e_{\text{nominal}}$  & 0.005 & constant & 5 \\
$q_{\text{tidal,nominal}}$  & 8400 & W m$^{-2}$ & \textit{this work} \\
$Q'$ & 1000 & constant & 6 \\
$k_2$ & 0.299 & constant & 7, 8 \\
$A$ & 0.1 & constant & 9 \\
$T_{\text{mantle},i}$ & 4000 & K & 11, 12 \\
$\phi_{\text{melt}}^{\text{crit}}$ & 0.4 & constant & 14 \\
$b_{\text{H},\text{O}}$ & $4.8 \times 10^{19} T_{\text{esc}}^{0.75}$ & $\text{m}^{-1} \text{ s}^{-1}$ & 15 \\
$b_{\text{H},\text{CO}_2}$ & $8.4 \times 10^{19}T_{\text{esc}}^{0.6}$ & $\text{m}^{-1} \text{ s}^{-1}$ & 15 \\
${C_{\text{U,total}}}$ & 21 & $\text{ppb}$ & 16 \\
$C_{{}^{238}\mathrm{U}}$ & $0.9927 \times {C_{\text{U,total}}}$ & $\text{ppb}$ & 16 \\
$C_{{}^{235}\mathrm{U}}$ & $0.0072 \times {C_{\text{U,total}}}$ & $\text{ppb}$ & 16 \\
$C_{{}^{232}\mathrm{Th}}$ & $4.01 \times {C_{\text{U,total}}}$ & $\text{ppb}$ & 16 \\
$C_{{}^{40}\mathrm{K}}$ & $1.28 \times {C_{\text{U,total}}}$ & $\text{ppb}$ & 16 \\
$H_{{}^{238}\mathrm{U}}$ & $9.37\times10^{-5}$ & $\text{W kg}^{-1}$ & 16 \\
$H_{{}^{235}\mathrm{U}}$ & $5.69\times10^{-4}$ & $\text{W kg}^{-1}$ & 16 \\
$H_{{}^{232}\mathrm{Th}}$ & $2.69\times10^{-5}$ & $\text{W kg}^{-1}$ & 16 \\
$H_{{}^{40}\mathrm{K}}$ & $2.79\times10^{-5}$ & $\text{W kg}^{-1}$ & 16 \\
$\lambda_{{}^{238}\mathrm{U}}$ & $0.155\times10^{-9}$ & $\text{yr}^{-1}$ & 17 \\
$\lambda_{{}^{235}\mathrm{U}}$ & $0.985\times10^{-9}$ & $\text{yr}^{-1}$ & 17 \\
$\lambda_{{}^{232}\mathrm{Th}}$ & $0.0495\times10^{-9}$ & $\text{yr}^{-1}$ & 17 \\
$\lambda_{{}^{40}\mathrm{K}}$ & $0.555\times10^{-9}$ & $\text{yr}^{-1}$ & 17 \\
$\alpha$ & $2 \times 10^{-5}$ & $\text{ K}^{-1}$ & 18 \\
$A$ & $2.4 \times 10^{-4}$ & $\text{Pa s}$ & 14, 18 \\
$B$ & $4600$ & $\text{K}$ & 14, 18 \\
$C$ & $3.8 \times 10^{9}$ & $\text{Pa s}$ & 14, 18 \\
$D$ & $3.5 \times 10^{5}$ & $\text{J mol}^{-1}$ & 14, 18 \\
$\alpha_n$  & $26$ & constant & 20 \\
$\beta$ & $1/3$ & constant & 14 \\
$C_0$ & $0.089$ & constant & 14 \\
$k_{\text{H}_2\text{O}}$ & $0.01$ & constant & 15 \\
$k_{\text{perov,CO}_\text{2}}$ & $5\times10^{-4}$ & constant & 15 \\
$k_{\text{lherz,CO}_\text{2}}$ & $2.1\times10^{-3}$ & constant & 15 \\
$R_\text{lhez-perov}$ & $390$ & km & 15 \\
\enddata
\tablecomments{1. \citet{hu_secondary_2024}; 2. \citet{crida_mass_2018}; 3. \citet{mamajek_improved_2008}; 4. \citet{meier_tobias_g_interior_2023}; 5. \citet{bolmont_emeline_tidal_2013}; 6. \citet{tobie_g_tidal_2019}; 7. \citet{yoder_astrometric_1995}; 8. \citet{henning_tidally_2009}; 9. \citet{essack_low-albedo_2020}; 10. \citet{bourrier_v_55_2018}; 11. \citet{fiquet_melting_2010}; 12. \citet{andrault_solidus_2011}; 13. \citet{howe_mass-radius_2014}; 14. \citet{lebrun_thermal_2013}; 15. \citet{zahnle_mass_1986}; 16. \citet{schubert_mantle_2001}; 17. \citet{mcnamara_cooling_2000}; 18 \citet{schaefer_predictions_2016}; 19. \citet{henning_tidally_2009}; 20. \citet{mei_influence_2002}.}
\end{deluxetable}

\subsection{Initial Volatile Content and Partitioning} \label{subsec:initialvolatiles}

We model the evolution of an outgassed $\text{CO}_2$ and $\text{H}_2\text{O}$ atmosphere on 55 Cnc e with a volatile evolution model based on the solubility of $\text{CO}_2$ and $\text{H}_2\text{O}$ in the magma ocean. The solubilities of volatiles in silicate melts have been demonstrated to depend on melt composition and oxidation state \citep{holloway_high-pressure_1992, papale_modeling_1997, kadik_formation_2004, morizet_coh_2010, mysen_solubility_2011, hirschmann_solubility_2012, dasgupta_carbon_2013, wetzel_degassing_2013, sossi_redox_2020}. However, solubilities have largely been measured in partial melting products such as basaltic or andesitic compositions. For comparison, Earth-like rocky mantles are thought to be primarily ultramafic in composition, such as peridotite and pyroxene-rich rock \citep{putirka_composition_2019, putirka_polluted_2021}. Although combined solubilities of CO$_2$ and H$_2$O have been measured in basalts, combined solubilities have not been measured in ultramafic compositions. H$_2$O has been experimentally measured in peridotite individually \citep{sossi_solubility_2023} and basaltic melts \citep{dixon_experimental_1995, newman_volatilecalc_2002, shishkina_solubility_2010} at low pressures (typically 0.1-5 GPa).

However at higher pressures characteristic of deep mantle melts, complete miscibility of H$_2$O is possible in both peridotite and basaltic melts, at pressures above 3.6 GPa and 5 GPa, respectively \citep{kessel_waterbasalt_2005, mibe_second_2007}. CO$_2$ solubility has also been measured at low pressures in the 0.1-5 GPa range \citep{pan_pressure_1991, dixon_experimental_1995, mysen_solubility_2011, wetzel_degassing_2013, stanley_solubility_2014, armstrong_speciation_2015} for both felsic and mafic melts, but is significantly lower than H$_2$O \citep{blank_solubilities_1993, fegley_volatile_2020, sossi_solubility_2023}. Because 55 Cnc e is often classified as a sub-Neptune or rocky super-Earth, the initial volatile reservoir may be significantly larger than Earth-sized planets, while deep-mantle pressures may enhance volatile solubility and miscibility. For reference, Earth's CMB is $\sim$135 GPa, while 55 Cnc'e may exceed $\sim$1200 GPa \citep{brooker_structural_2001, meier_tobias_g_interior_2023}. However, mantle-atmosphere interactions are limited by the solubility at the surface of the magma ocean, which may result in a sub-saturated volatile reservoir in the deeper mantle. At certain temperature-pressure and redox conditions, carbon saturation in the silicate melt can lead to the exsolution of graphite, diamond or other C-bearing species \citep{hirschmann_magma_2012, madhusudhan_possible_2012}; however we do not consider carbon depletion through exsolution of these secondary phases, but rather assume available dissolved carbon is readily outgassed. With our models, we produced simulations with up to 10 wt\% H$_2$O and CO$_2$ each, up to a combined mass fraction of 20 wt\% for total dissolved volatiles.

We first model the volatile mass balance as the distribution of volatiles between the solid crystallizing mantle, magma ocean, and atmosphere system \citep{elkins-tanton_linked_2008, lebrun_thermal_2013, schaefer_predictions_2016}, which is parameterized as:

\begin{equation}\label{eqn:mass_Balance}
\begin{aligned}
   M_{i}^{\text{total}} &= M_{i}^{\text{sol}} + M_{i}^{\text{liq}} + M_{i}^{\text{atm}} \\
    &= k_{i} F_{i}^{\text{liq}} M_{i}^{\text{sol}} + F_{i}^{\text{liq}} M_{i}^{\text{liq}} + \frac{4\pi R_p^2}{g} P_i
\end{aligned}
\end{equation}

\noindent where $i$ is the volatile species $\text{H}_2\text{O}$ or $\text{CO}_2$, $k_i$ is the solid-liquid partition coefficient for each species, $F_{i}^{\text{liq}}$ is the mass fraction of volatile $i$ in the silicate melt, and $P_i$ is the partial pressure of the volatile in the atmosphere. The atmospheric pressure of each volatile can be found based on empirical solubility relationships developed by \citet{papale_modeling_1997} for water, and \citet{pan_pressure_1991} for $\text{CO}_2$, from the relations:

\begin{equation}\label{eqn:PH2O_Papale}
P_{\text{H}_2\text{O}} = \left( \frac{F_{\text{H}_2\text{O}}^{\text{liq}}}{3.44 \times 10^{-8}} \right)^{1/0.74}
\end{equation}

\begin{equation}\label{eqn:PCO2_Pan}
P_{\text{CO}_2} = \frac{F_{\text{CO}_2}^{\text{liq}}}{4.4 \times 10^{-12}}
\end{equation}

\noindent where the volatile mass fraction in melt $F_i^{\text{liq}}$, is the same as Eq. (\ref{eqn:mass_Balance}). For the dimensionless $k_{\text{H}_2\text{O}}$, we use an approximation of $k_{\text{H}_2\text{O}}=0.01$ from \citet{schaefer_predictions_2016}, though values for $k_{\text{H}_2\text{O}}$ for different mineral/melt pairs ranges from  $10^{-4}$ to $1.1 \times 10^{-2}$ \citep{elkins-tanton_linked_2008, lebrun_thermal_2013}. For $k_{\text{CO}_2}$, we use the liquid-solid partition equation from Lebrun et al. (2013) using the mass of the solid crystallized mantle, which results in the combined parameterization for both volatiles:

\begin{equation}\label{eqn:kh2O_partition}
k_{\text{H}_2\text{O}}=0.01
\end{equation}

\begin{equation}\label{eqn:kCO2_partition}
k_{\text{CO}_2} = \frac{M_{\text{perov}} k_{\text{perov,CO}_\text{2}} + M_{\text{lherz}} k_{\text{lherz,CO}_\text{2}}}{M_{\text{perov}} + M_{\text{lherz}}}
\end{equation}

\noindent where the partition coefficients for $\text{CO}_2$ in perovskite and lherzolite are given by $k_{\text{perov,CO}_\text{2}}$ and $k_{\text{lherz,CO}_\text{2}}$ from \citet{lebrun_thermal_2013}. The fractional solid phase perovskite and lherzolite components, $M_{\text{perov}}$ and $M_{\text{lherz}}$, are determined by a lherzolite-perovskite boundary at 23 GPa \citep{lebrun_thermal_2013, takahashi_melting_1993},

\begin{equation}\label{eqn:M_perov}
M_{\text{perov}} = \rho \frac{4}{3}\pi \Big[ (R_p - R_{\text{lhez-perov}})^3 - R_c^3 \Big]
\end{equation}

\begin{equation}\label{eqn:M_lher}
M_{\text{lherz}} = \rho \frac{4}{3}\pi \Big[ R_p^3 - (R_p - R_{\text{lhez-perov}})^3 \Big]
\end{equation}

\noindent where the masses of each component are approximated as spherical shells defined by the planetary radius $R_p$, core radius $R_c$, and approximate lherzolite-perovskite boundary depth for rocky planets, at 23 GPa, to be $R_\text{lhez-perov}=390\text{ km}$.

Solving for the mass fraction of volatiles in melt $F_i^{\text{liq}}$ from Eq. (\ref{eqn:mass_Balance}), we use a series of differential equations to model the evolution of sequestered CO$_2$ and H$_2$O in the solid mantle, and combined magma ocean plus atmosphere system,

\begin{equation}\label{eqn:volatile_sol_evol}
\frac{dM_{i}^{\text{sol}}}{dt} = \frac{dr_s}{dt} \ k_i F_{i}^{\text{liq}} 4 \pi \rho_m r_s^2
\end{equation}

\begin{equation}\label{eqn:volatile_liq_evol}
\frac{dM_{i}^{\text{liq+atm}}}{dt} = 
\begin{cases}
-\frac{dM_{\text{H}_2\text{O}}^{\text{sol}}}{dt} - 4\pi R_p^2 \Phi_{\text{H}} \frac{\mu_{\text{H}_2\text{O}}}{2\mu_{\text{H}}} &, i =\text{H}_2\text{O}\\
-\frac{dM_{\text{CO}_2}^{\text{sol}}}{dt} -  4\pi R_p^2 \Phi_{\text{CO}_2}  \mu_{\text{CO}_2} &, i =\text{CO}_2\\
\end{cases}
\end{equation}

Our volatile evolution model primarily depends on the rate of solidification of the magma ocean ($\frac{dr_s}{dt}$), and the efficiency of hydrodynamic-driven loss, characterized by escape rates $\Phi_{i}$ for the respective escaping species relevant to $\text{H}_2\text{O}$ and $\text{CO}_2$ as described in Eq. (\ref{eqn:mass_flux}).

For oxygen, we follow the same assumptions as \citet{schaefer_predictions_2016}, where the formation of oxygen in the atmosphere is produced from the photodissociation of $\text{H}_2\text{O}$, at the same time governed by an equilibrium with iron species in the liquid melt. We use a magma ocean redox buffering model in which atmospheric oxygen reacts with $\text{FeO}$ in silicate melt, leading to the creation of $\text{Fe}_{2}\text{O}_3$ in the liquid melt \citep{hirschmann_solubility_2012, deng_magma_2020, hirschmann_magma_2022},

\begin{equation}\label{eqn:redox_buffer}
\text{FeO} \ (\text{liq}) +\frac{1}{4}\text{O}_2 \ (\text{atm}) =\text{FeO}_{1.5} \ (\text{liq})
\end{equation}

\noindent We use the empirically-derived equation relating oxygen fugacity $f_{\text{O}_2}$ to the ratio of $\text{Fe}^{3+}$ and $\text{Fe}^{2+}$ in the melt from \citet{kress_compressibility_1991}. The relation adopts an empirical calibration of the non-ideal activity coefficients based on silicate melt composition (in mole fractions of the major oxide components ${X}_{i}$) to determine the $\text{Fe}^{3+}/\text{Fe}^{2+}$ ratio as a function of temperature and pressure:

\begin{equation}\label{eqn:oxygen_fugacity}
\begin{aligned}
\log \left( \frac{\text{Fe}^{3+}}{\text{Fe}^{2+}} \right) &=
-1.828 X_{\mathrm{FeO}}
+ 3.201 X_{\mathrm{CaO}} \\ 
&
+ 5.854 X_{\mathrm{Na_2O}}
+ 6.215 X_{\mathrm{K_2O}}
- 2.243 X_{\mathrm{Al}_2\mathrm{O}_3}\\ 
&
+ \frac{11492}{T}
- \frac{7.01 \times 10^{-7} P}{T}\\ 
&
- \frac{1.54 \times 10^{-10} P (T - 1673)}{T}\\ 
&
+ \frac{3.85 \times 10^{-17} P^2}{T}\\ 
&
- 3.36 \left( 1 - \frac{1673}{T} - \ln \left( \frac{T}{1673} \right) \right)
- 6.675
\end{aligned}
\end{equation}

\noindent We assume the composition of the silicate melt is that of the Bulk Silicate Earth (BSE). We choose the initial mass fraction of total $\text{FeO}$ in the mantle to be 0.08 ($\text{FeO}=8\text{ wt\%}$), assuming all $\text{Fe}$ in the mantle to be in the form of $\text{FeO}$, with 2\% of total FeO being sequestered in $\text{Fe}_2\text{O}_3$ initially, an approximation for the lower end concentration of $\text{Fe}_2\text{O}_3$ in the mantle source of mid-ocean ridge basalts \citep{mallmann_crystalmelt_2009, sorbadere_behaviour_2018, davis_partitioning_2021}.

We follow the parameterization of \citet{schaefer_predictions_2016} for the evolution of atomic $\text{O}$ in the solid mantle and liquid melt, with equations analogous to $\text{H}_2{\text{O}}$ (Eq. (\ref{eqn:volatile_sol_evol})) and $\text{CO}_2$ (Eq. (\ref{eqn:volatile_liq_evol})), using the same assumption that fractionation between liquid melt and crystallized mantle for both $\text{FeO}$ and $\text{FeO}_{1.5}$ is minimal (although see \citet{schaefer2024ferric}), yielding:

\begin{equation}\label{eqn:oxygen_sol_evol}
\frac{dM_{\text{O}}^{\text{sol}}}{dt} = \frac{d r_s}{dt} F_{\text{FeO}_{1.5}}^{\text{liq}} 4\pi \rho_m r_s^2 \frac{\mu_{\mathrm{O}}}{2\mu_{\mathrm{FeO}_{1.5}}}
\end{equation}

\begin{equation}\label{eqn:oxygen_liq_evol}
\frac{dM_{\text{O}}^{\text{liq+atm}}}{dt}  = - \frac{dM_{\text{O}}^{\text{sol}}}{dt} +
4 \pi R_p^2 \left( 
\Phi_{\text{H}}  \frac{\mu_{\text{O}}}{2 \mu_{\text{H}}}
- \Phi_{\text{O}} \right)
\end{equation}

\noindent where $F_{\text{FeO}_{1.5}}^{\text{liq}}$ is the mass fraction of $\text{Fe}\text{O}_{1.5}$ in the liquid melt of the magma ocean. Oxygen abundance evolution depends on: (1.) sequestration by ferrous oxide in the mantle, (2.) replenishment by XUV-driven dissociation of $\text{H}_2\text{O}$, and (3.) hydrodynamic drag of oxygen. Because our volatile model only considers the evolution of $\text{H}_2\text{O}$, $\text{CO}_2$, and $\text{O}_2$, we discuss the ramifications of a reduced interior producing an outgassed $\text{CO}$ atmosphere in the place of $\text{CO}_2$ in Section \ref{sec:discussion_reduced_mantle}. For purposes of modeling an evolved $\text{CO}_2$-dominated atmosphere on 55 Cnc e, Eq. (\ref{eqn:volatile_sol_evol}), (\ref{eqn:volatile_liq_evol}), (\ref{eqn:oxygen_sol_evol}), and (\ref{eqn:oxygen_liq_evol}) describe our complete volatile model as an expansion of the interior-atmosphere model employed by \citet{schaefer_predictions_2016} for magma ocean worlds.

\subsection{Tidal Heating Model}\label{subsec:tidal_model}

Heat deposition by tidal forces is suggested to be an important component of rocky exoplanet evolution due to its impact on volcanism, atmospheric development, orbital dynamics, and geophysical properties \citep{jackson_tidal_2008, driscoll_tidal_2015, dobos_vera_tidal_2019, farhat_tides_2025}. \citet{mcintyre_s_r_n_tidally_2022} calculated possible tidal heating rates for 767 tidally-locked rocky exoplanets and candidates, suggesting tidal heating may be a significant contributor to raising planetary temperatures. \citet{farhat_tides_2025} improved on existing tidal heating models for magma ocean worlds by adding liquid viscoelastic responses to tidal forces, parameterizing the Love number $k_2$ for liquid-state tidal heating models. Though integration of liquid tides is beyond the scope of this paper, we discuss the implications of a complete liquid-solid tidal theory on eccentricity and heat flux evolution in Section \ref{sec:discussion_tidal_model}.

We include tidal heating as an additional input to our planetary thermal evolution equation (Eq. \ref{eqn:mantle_temp}), which supplements the secular heat of formation and radioactive decay $Q_{\text{rad}}$. We adopt four fiducial eccentricity values, for which we determine the amount of tidal heating that would have been produced during tidal circularization and orbital evolution. An assumed present-day value of $e=0.001$ for the eccentricity was used, with a realistic maximum value estimated to be $e < 0.015$ and a plausible value of $e=0.005$ or less, following \citet{bolmont_emeline_tidal_2013}, though observations with RV and transit data have found eccentricities to be potentially higher \citep{nelson_55_2014, baluev_enhanced_2015, demory_map_2016, bourrier_v_55_2018, mello_tidal_2025}. However, circularization timescales for 55 Cnc e to reach near-zero eccentricity in synchronized rotation have been shown to be $< 3$ Myr \citep{mello_tidal_2025}, with pumping by companion planets in the system maintaining the observed non-zero eccentricity. Table \ref{tab:eccentricity_val} lists the range of reported eccentricity values for 55 Cnc e.

\begin{deluxetable*}{lccc}
\center
\tablecaption{Reported Observed or Modeled Eccentricities for 55 Cnc e 
\label{tab:eccentricity_val}}
\tablewidth{1\columnwidth}
\tablehead{
\colhead{\textit{e}} & \colhead{Observation or Model Type} & Instrument & \colhead{Reference}
}
\startdata
0.00498 & Model (\textit{N}-body) & -- & \citet{bolmont_emeline_tidal_2013} \\
$<0.018$ & Model (synchronous, `hard-body') & -- & \citet{mello_tidal_2025} \\
$0.028^{+0.022}_{-0.019}$ &  RV (`Case~2') & Lick, Keck, HET, Harlan J. Smith &  \citet{nelson_55_2014} \\
$<0.034$ & Model (synchronous, `soft-body') & -- & \citet{mello_tidal_2025} \\
$0.034^{+0.022}_{-0.021}$ & RV (`Case~1') & Lick, Keck, HET, Harlan J. Smith & \citet{nelson_55_2014} \\
0.040 & RV (\citet{nelson_55_2014} Reanalysis) & Lick, Keck, HET/HRS, Harlan J. Smith & \citet{baluev_enhanced_2015} \\
$0.05^{+0.03}_{-0.03}$ & RV & Lick, Keck, HET/HRS, HARPS, SOPHIE  & \citet{bourrier_v_55_2018} \\
$0.142^{+0.06}_{-0.066}$ & RV (\citet{mcarthur_detection_2004} Reanalysis) & HET/HRS & \citet{rosenthal_california_2021} \\
$<0.06$ & Secondary Eclipse & Spitzer/IRAC  & \citet{demory_detection_2012} \\
$<0.19$ & Transit, Secondary Eclipse & Spitzer/IRAC &\citet{demory_variability_2016} \\
\enddata
\tablecomments{Eccentricities are given as ascending nominal mean values, unless errors are reported.}
\end{deluxetable*}

The tidal heating rate from \citet{driscoll_tidal_2015} for synchronous rotation is modeled as:

\begin{equation}\label{eqn:tidal_heating}
q_{\text{tidal}} = -\frac{21}{2} \operatorname{Im}(k_2) G^\frac{3}{2} M_*^\frac{5}{2} R_p^5 \frac{e^2}{a^\frac{15}{2}}
\end{equation}

\noindent where $k_2$ is the Love number of the planet. The Love number $k_2$ and tidal response factor $Q$ represent rheological response parameters from tidal forces, for which we use an Earth-like parametrization $k_2=0.299$ \citep{yoder_astrometric_1995, henning_tidally_2009}, and an order of magnitude super-Earth/sub-Neptune approximation of $Q' =1000$, between rough values for Earth-like ($Q' =10^2$) to Neptune-like ($Q' =10^4$) $Q'$ estimates \citep{tobie_g_tidal_2019}. From \citet{peale_contribution_1978}, the resulting Q and $\text{Im}(k_2)$ values are derived from the following rheological relations:

\begin{equation}\label{eqn:Q_tidal_factor}
Q = Q'  \frac{2 k_2}{3}
\end{equation}

\noindent and,

\begin{equation}\label{eqn:Q'_tidal_factor}
\text{Im}(k_2) = \frac{k_2}{Q}.
\end{equation}

\noindent Using the tidal heating potential $q_{\text{tidal}}$, the tidally-enhanced equilibrium temperature $(T_{\text{eq}})$ can be determined analytically with the following equation from \citet{reinhold_ignan_2025},

\begin{equation}\label{eqn:tidal_Teq}
T_{\text{eq}}^{\text{tidal}} = \left( \frac{S_{\text{flux}}+q_{\text{tidal}}}{\sigma} \right)^{\frac{1}{4}}
\end{equation}

\noindent where $S_{\text{flux}}$ $\text{(W m}^{-2})$ is the effective global planetary insolation after accounting for albedo and hemispherical correction ($S_{\text{flux}}=(1-A)L_{\star}/16 \pi a^2$), and $\sigma$ is the Stefan–Boltzmann constant. We use this analytical expression in place of our numerical model (Eq. \ref{eqn:surf_temp}) for an approximation of the surface temperature for a bare rocky surface based on tidal heating and insolation alone. We use this temperature to gauge which exoplanets may harbor magma oceans and tidally-driven outgassing for further study (Fig. \ref{fig:Mass_Radius_Magma_Ocean}). 


\section{Results}\label{sec:results}

To contextualize our results, we summarize the processes that control volatile outgassing in the magma ocean stage below. Initial outgassing occurs to set the equilibrium between the atmosphere and dissolved volatiles, typically lasting  $<1$ Myr. Minor outgassing continues as atmospheric escape drives the system away from the atmospheric pressure corresponding to equilibrium with the melt. This process persists throughout the lifetime of the magma ocean. Additional outgassing occurs as the deep mantle solidifies, since volatiles are less soluble in solid phases and therefore concentrate in the residual liquid at values above the solubility saturation level. As the melt becomes enriched in volatiles, outgassing acts to re-establish a Henry's law-type equilibrium. Solidification that leads to outgassing is triggered by a decrease in surface temperature. For a planet of a given size, there exists a critical surface temperature at which this occurs, with outgassing continuing under these conditions until the planet either reaches thermal equilibrium or becomes fully solidified. In Section \ref{sec:results_tidal_greenhouse}, we examine the resulting long-term outgassing behavior that emerges from these interactions in more detail.

As we will show below, our model demonstrates that without the addition of tidal heating, thermal evolution and outgassing in the first few Myr are primarily dependent on initial concentrations of CO$_2$ and H$_2$O because of the strong greenhouse effect on surface temperature, demonstrating that a higher initial volatile reservoir can prolong outgassing while producing a longer-lived and deeper magma ocean in the early planetary history (Fig. \ref{fig:WMFvsFCO2_temp} and \ref{fig:TeqTime}). The present-day atmospheric composition of 55 Cnc e can be used to infer the size of the initial planetary volatile inventory of $\text{H}_2\text{O}$ and $\text{CO}_2$. We discuss these results in Section \ref{sec:results_greenhouse}.

With tidal heating, tenuous outgassing occurs due to prolonged cooling on Myr-Gyr timsecales. Because of the extreme orbit and close proximity of 55 Cnc e to its host star, eccentricity-driven tidal heating contributes to a substantial increase in the planet's temperature and can influence the outgassing of $\text{H}_2\text{O}$ and $\text{CO}_2$ to the present-day. However, outgassing rates are ultimately determined by the combined effect between the mantle volatile reservoir size and the eccentricity, with larger eccentricities driving higher tidal heating rates.  Continued outgassing depends on both the delayed cooling due to the input of tidal heating and the availability of volatiles to outgas. We make predictions for 55 Cnc e's observable atmosphere size ($P_{\text{atm}}<1000$ bars) based on model simulations of possible eccentricity and reservoir sizes. These conclusions are examined in more detail in Sections \ref{sec:results_tidal_outgassing} and \ref{sec:results_tidal_greenhouse}.

\subsection{The Greenhouse Effect on Atmospheric Composition and Early Outgassing}\label{sec:results_greenhouse}

\begin{figure*}
\center 
\includegraphics[width = 2.15\columnwidth]{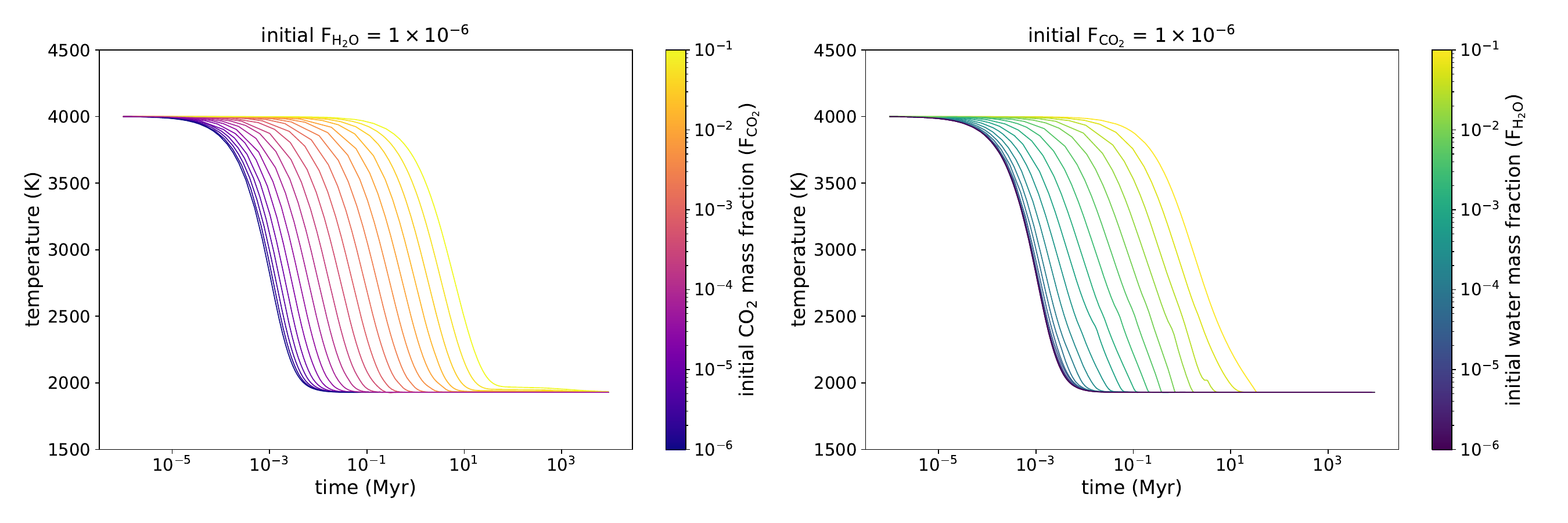}
\caption{ {\bf Left:} Surface temperature evolution of 55 Cnc e with varying initial $\text{CO}_2$ mass fractions $F_{\text{CO}_2}$ at a constant initial $\text{H}_2\text{O}$ mass fraction ($F_{\text{H}_2\text{O}}$) of $1\times10^{-6}$. {\bf Right:} Surface temperature evolution with variable initial $F_{\text{H}_2\text{O}}$, at a fixed initial $F_{\text{CO}_2}$ of $1\times10^{-6}$. Both sets of simulations assume no tidal heating. The highest starting abundances of $\text{H}_2\text{O}$ or $\text{CO}_2$ can delay cooling of the magma ocean because of the greenhouse effect.}
\label{fig:WMFvsFCO2_temp}
\end{figure*}

\begin{figure}
\center 
\includegraphics[width = 1.\columnwidth]{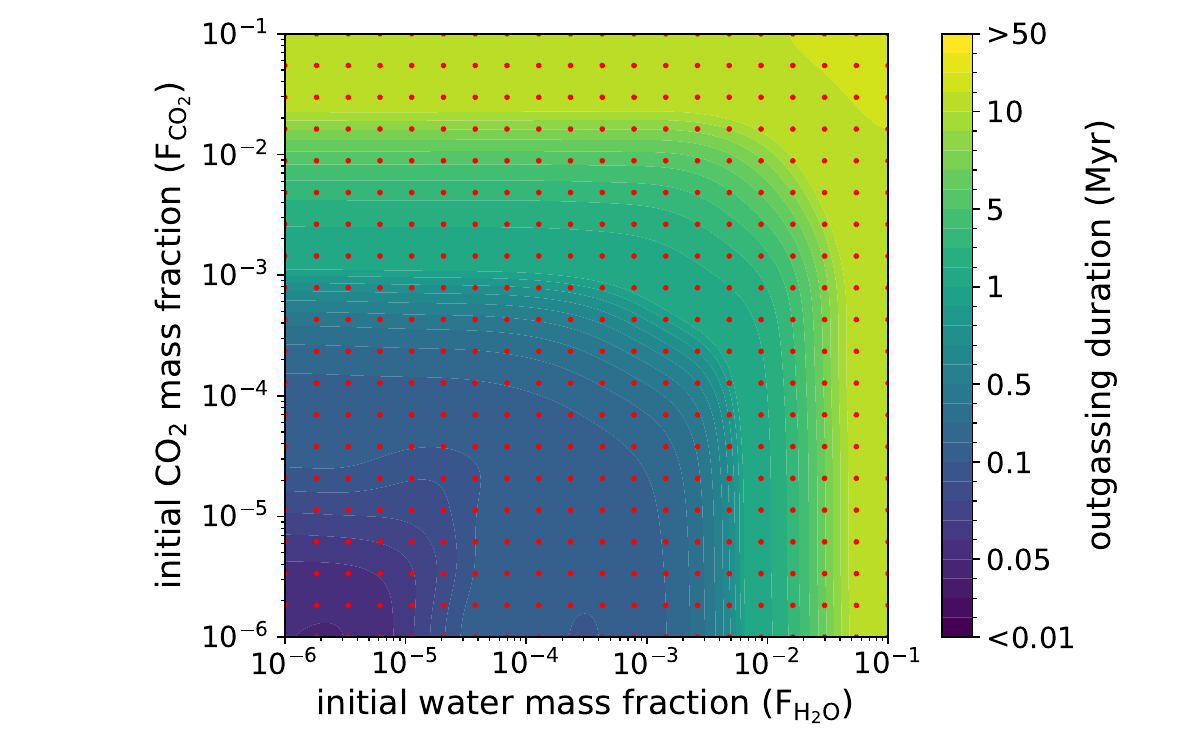}
\caption{Outgassing durations for a combination of initial H$_2$O and CO$_2$ mass fractions, with no tidal heating assumed. The cooling of the magma ocean and subsequent outgassing can be prolonged up to $\sim$30 Myr with the highest initial volatile concentrations as a result of greenhouse warming. Red markers denote the 400 total computed simulation grid points. The bottom and left axes lines are equivalent to simulations plotted in Fig. \ref{fig:WMFvsFCO2_temp} for respective initial $F_{\text{H}_2\text{O}}$ and $F_{\text{CO}_2}$ values.}
\label{fig:TeqTime}
\end{figure}

\begin{figure*}
\center 
\includegraphics[width = 2.15\columnwidth]{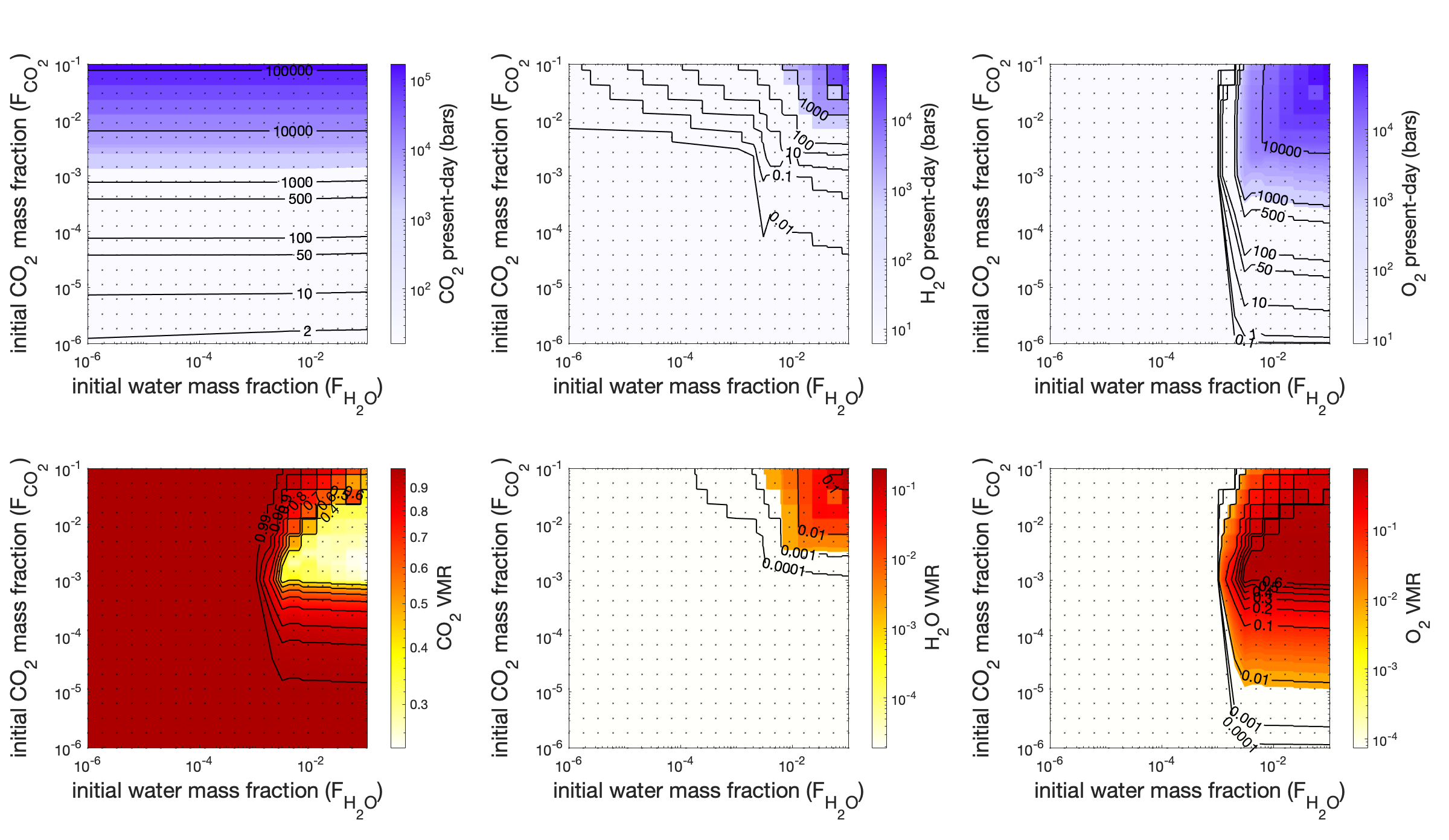}
\caption{ Grid of simulations of atmospheric pressure contribution (bars) and total volume mixing ratio (VMR) for each molecular species as a result of initial volatile abundances. Results are shown for the current age of the system and assume no tidal heating ($e=0$). Small black dots denote the simulation grid points. {\bf Top Left:} CO$_2$ atmospheric pressure (bars). {\bf Top Center:} H$_2$O atmospheric pressure (bars). {\bf Top Right:} O$_2$ atmospheric pressure (bars). {\bf Bottom Left:} CO$_2$ VMR. {\bf Bottom Center:} H$_2$O VMR. {\bf Bottom Right:} O$_2$ VMR. }
\label{fig:WMFvsFCO2}
\end{figure*}

For models without tidal heating, the greenhouse effect regulates the cooling and solidification of the magma ocean. We explored initial volatile mass fractions from $10^{-4}$ to 10 \text{wt\%} ($10^{-6} \text{ to } 10^{-1}$ $F_{\text{H}_2\text{O}}$ or $F_{\text{CO}_2}$) for both species, as discussed in Section \ref{subsec:initialvolatiles}. Figure \ref{fig:WMFvsFCO2_temp} shows the thermal evolution of the surface temperature on 55 Cnc e for different initial volatile concentrations, with the mass fractions of either $F_{\text{CO}_2}$ (right) or $F_{\text{H}_2\text{O}}$ (left) held constant at the minimum value for each species ($1\times10^{-6}$). The time for the surface temperature (shown in Fig. \ref{fig:WMFvsFCO2}) to cool to steady state dictates the duration of the strongest and earliest outgassing period of the secondary atmosphere, which is due to the initial volatile equilibrium between the atmosphere and magma ocean (Fig. \ref{fig:TeqTime}). The evolved pressures of atmospheric species and their subsequent greenhouse warming effect play an important role in regulating the duration of outgassing in the short term, while influencing composition on the longer term, which is discussed in the following paragraph and figure (Fig. \ref{fig:WMFvsFCO2}).

The duration of outgassing for all 400 simulations with variable starting volatile inventories is presented in Figure \ref{fig:TeqTime}. At most, an extension of $\sim$30 Myr of outgassing is possible, with lower volatile abundances suggesting a $<1$ Myr period of strong outgassing on a young 55 Cnc e. The duration of outgassing depends more strongly on the initial $F_{\text{CO}_2}$, which is demonstrated by an asymmetry along the diagonal axis. This is because $\text{CO}_2$ remains in the atmosphere, unlike water vapor, which experiences XUV-driven escape. 

As shown in Figure \ref{fig:WMFvsFCO2}, we predict the atmospheric composition of 55 Cnc e to be $\text{CO}_2$-dominated, with a volume mixing ratio (VMR) $> 99\%$ for most simulations that result in a significant atmosphere ($> 2$ bar). However, even if the starting water abundance is high ($>0.1$ wt\%), a sufficient inventory of $\text{CO}_2$ is needed to outgas in order to preserve the unstable and lighter molecular weight species of $\text{H}_2\text{O}$ and $\text{O}_2$ from being lost quickly in the early planetary history. In order to retain a minimum 1\% VMR of $\text{O}_2$ or $\text{H}_2\text{O}$ at the present-day, an initial $F_{\text{CO}_2}$ of $\sim$0.001 wt\% or $\sim$0.1 wt\% is required, respectively, as seen in in Figure \ref{fig:WMFvsFCO2}.

Note that our model does not include CO, but we discuss the possibility for $\text{CO}$ formation and outgassing from a reduced mantle in Sections \ref{sec:discussion_photochemistry} and \ref{sec:discussion_reduced_mantle}.

\subsection{Tidal Heating Can Further Prolong Outgassing of $\text{CO}_2$ on Myr-Gyr Timescales}\label{sec:results_tidal_outgassing}

The addition of tidal heating can maintain an elevated surface temperature greatly exceeding that produced by stellar insolation alone, which results in prolonged cooling and outgassing. This can occur even at low eccentricities ($e<0.01$). Although initial volatile reservoirs can prolong outgassing on Myr timescales, without tidal heating, even very elevated initial volatile inventories cannot prolong outgassing of a secondary atmosphere to Gyr-timescales alone.

\begin{figure*}
\center 
\includegraphics[width = 2.15\columnwidth]{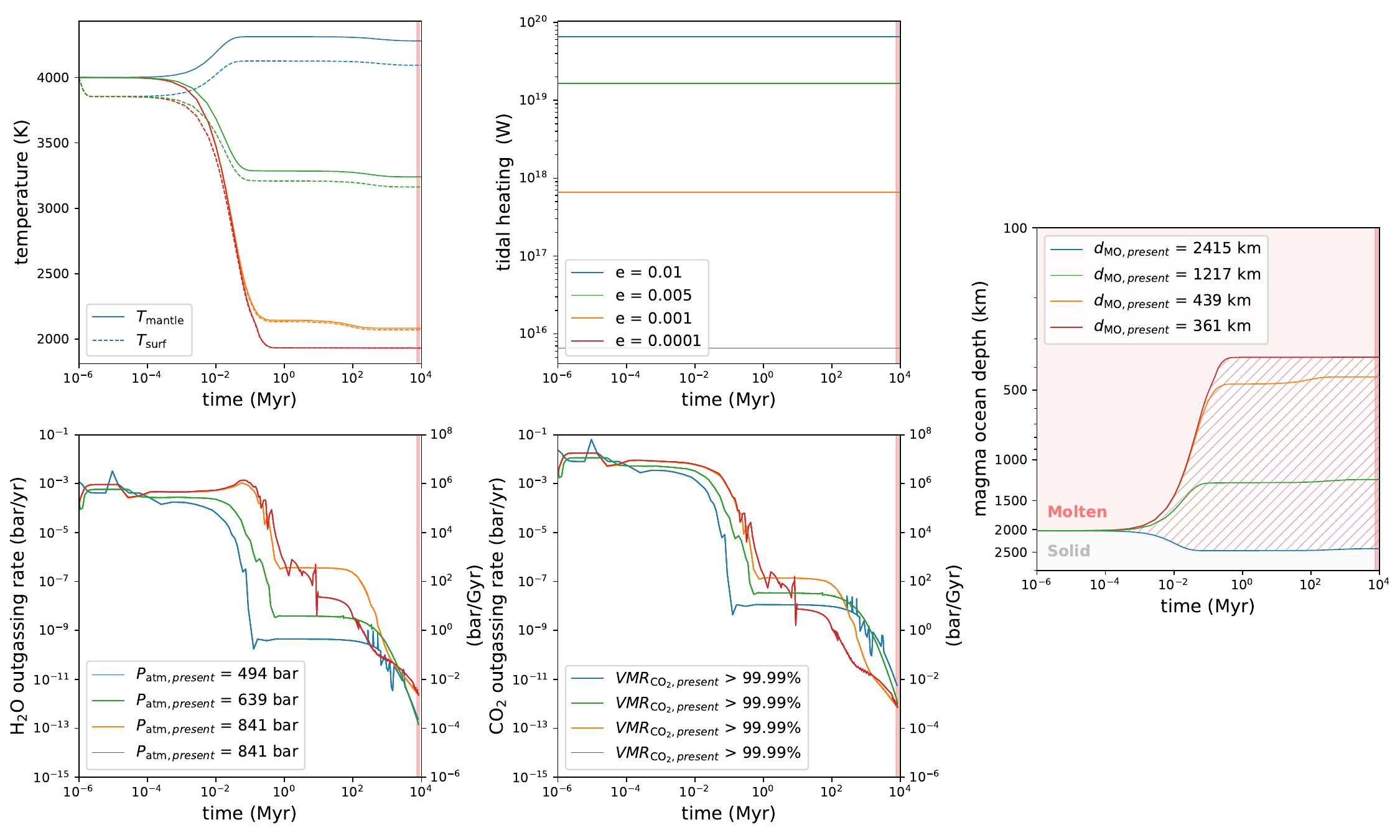}
\caption{ Thermal evolution and outgassing rates on 55 Cnc e through four tidal heating models. This figure uses an initial $F_{\text{H}_2\text{O}}$ of $5 \times 10^{-4}$ and initial $F_{\text{CO}_2}$ of $5 \times 10^{-4}$, and the total pressure of the atmospheres at the present-day are given by the legend in the lower left panel. The vertical red line marks the estimated age of the planet from \citet{mamajek_improved_2008}. Gaussian smoothing to reduce numerical noise was applied to $\text{H}_2\text{O}$ and $\text{CO}_2$ outgassing rates. Some residual noise persists, but large-scale trends are unchanged. {\bf Top Left:} Temperature evolution of surface (dotted) and mantle (solid)  (K). {\bf Top Center:} Respective orbital eccentricity tracks reflecting the various tidal heating contributions to the planetary interior. Note that the no tidal heating case ($e=0$) is analogous to the evolution of our lowest tidal heating case ($e=0.0001$) due to the extreme insolation making tidal heating negligible. We include an improbable high $e=0.01$ (blue) tidal heating case to show an upper ceiling on realistic eccentricities, where tidal forcing becomes so strong that the planet continues to heat rather than cool. {\bf Bottom Left:} $\text{H}_2\text{O}$ outgassing rates from magma ocean formation to present-day (bar/yr). {\bf Bottom Center:} CO$_2$ outgassing rates from magma ocean formation to present-day (bar/yr). Note that the predicted present-day compositional volume mixing ratio (VMR) of $\text{CO}_2$ is labeled in this panel as well. \textbf{Right:} Globally-averaged magma ocean depth (km) evolution throughout time. The red region marks parameter space that is molten in all simulations, while the white region marks parameter space that remains fully solid in all simulations. The hatched red region shows parameter space that may be molten or solid depending on the tidal heating model. This constrains globally-averaged present-day magma ocean depths from a shallower $d_{\text{MO}}=361$ km to a deeper $d_{\text{MO}}=2415$ km.}
\label{fig:Tidal_overview}
\end{figure*}

Figure \ref{fig:Tidal_overview} shows the evolution of four tidal heating models and their associated eccentricity tracks for 55 Cnc e, isolating the effect of eccentricity on outgassing by holding the initial volatile inventory fixed. The simulations use initial $F_{\text{H}_2\text{O}}$ and$F_{\text{CO}_2}$ of $5 \times 10^{-4}$, which produces an atmosphere less than 1000 bars. The model was run to the estimated present-day age of $\sim$8 Gyr \citep{mamajek_improved_2008} from the time of magma ocean formation, a proxy for the post-protoplanetary accretion stage. For this set-up, simulations with a significant tidal heating contribution ($e=0.01$, $e =0.005$), were able to maintain a heightened $\text{CO}_2$ outgassing rate greater than 1 bar/Gyr at 1 Gyr. The outgassing rate decreased to less than 0.01 bar/Gyr at the present day, but was still greater than simulations with a low tidal heating rate ($e=0.001$, $e=0.0001$). $\text{H}_2\text{O}$ outgassing rates were generally lower at both early and late times for models with higher eccentricity than those with low eccentricity, potentially suggesting a trade-off between volatile-type and subsequent outgassing rates due to volatile solubility. Additionally, low tidal heating models exhibited higher $\text{CO}_2$ and $\text{H}_2\text{O}$ outgassing rates during the first 1-500 Myr due to faster crystallization of the magma ocean and mantle. Figure \ref{fig:Tidal_overview} also illustrates the tradeoff between outgassing rates and atmospheric thickness, with higher eccentricity models prolonging late-stage outgassing but also limiting the maximum outgassed atmosphere possible. This is a result of a deeper and larger magma ocean, which can sequester available volatiles. For our high eccentricity models ($e=0.01$, $e =0.005$), resultant total atmospheric pressures for the present-day were $P_{\text{atm}}=494$ bars and $P_{\text{atm}}=639$ bars respectively, which are lower than the total pressures predicted for smaller eccentricities ($e=0.001$ and $e=0.0001$), which both yield the maximum value of $P_{\text{atm}}=841$ bars. Note that varying the initial mantle temperature $T_{\text{mantle},i}$ between $\sim$3500-4500 K is not expected to significantly affect early magma ocean cooling or associated outgassing timescales. Efficient thermal cooling drives the system towards a metastable thermal equilibrium within $\sim$1~Myr in most molten regimes, even with differing stellar and initial mantle temperature cases \citep{lebrun_thermal_2013, schaefer_predictions_2016}, making initial mantle temperature differences negligible compared to early greenhouse regulation.

\subsection{Tidally-Enhanced Outgassing Depends on the Combined Effect of Volatiles and Eccentricity}\label{sec:results_tidal_greenhouse}

\begin{figure*}
\center 
\includegraphics[width = 2.\columnwidth]{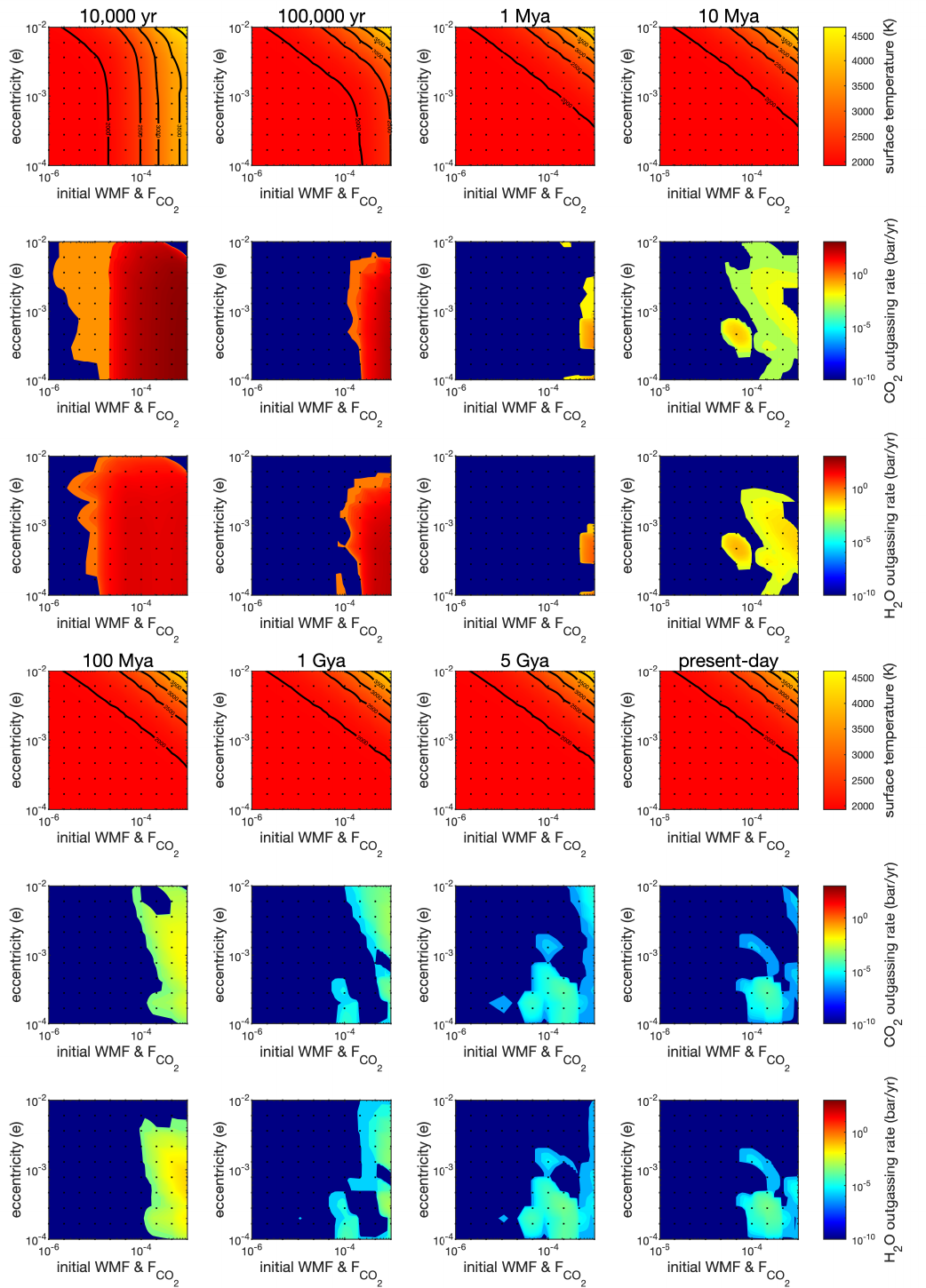}
\caption{Time series of surface temperature evolution, $\text{H}_2\text{O}$ outgassing rate, and $\text{CO}_2$ outgassing rate at eight time intervals as a function of eccentricity and initial equal parts $F_{\text{H}_2\text{O}}$ and $F_{\text{CO$_2$}}$ reservoirs. For simplicity, eccentricity tracks are kept at the constant initial value. Black markers denote the 100 total computed simulation grid points.}
\label{fig:Combined_Tidal_Grid}
\end{figure*}

Tidal enhancement of an outgassed secondary atmosphere was found to be a function of both volatile inventory size and eccentricity. Figure \ref{fig:Combined_Tidal_Grid} shows the evolution of surface temperature (K), and both $\text{H}_2\text{O}$ and $\text{CO}_2$ outgassing rates $(\text{bar yr}^{-1})$ as a time series from magma ocean formation to the present-day. Our model results demonstrate outgassing events can be broadly classified in three stages for a constant eccentricity evolution: (1.) The bulk of the atmosphere is formed as an initial large outgassing event $<$ 1 Myr during early cooling of the magma ocean, which also sets the surface temperature for the remainder of 55 Cnc e's lifetime. (2.) In the planet's midlife, between $\sim$1 Myr and $\sim$5 Gyr, higher outgassing rates are centered around $e=0.001$ at higher initial volatile abundances. Higher eccentricities do not correlate with higher outgassing rates. In general we find that outgassing has a relatively flat behavior, depending much more strongly on volatile abundance. (3.) At ages over 5 Gyr, evolution models with low eccentricities ($e < 0.001$) and moderately volatile-rich interiors ($F_{\text{H}_2\text{O}}$ and $F_{\text{CO}_2}$ > $\sim0.01$ wt\% each) allow further cooling and a lower surface temperature to be reached, which results in additional volatile outgassing if the interior has not yet fully desiccated.

\begin{figure}
\center 
\includegraphics[width = 1.\columnwidth]{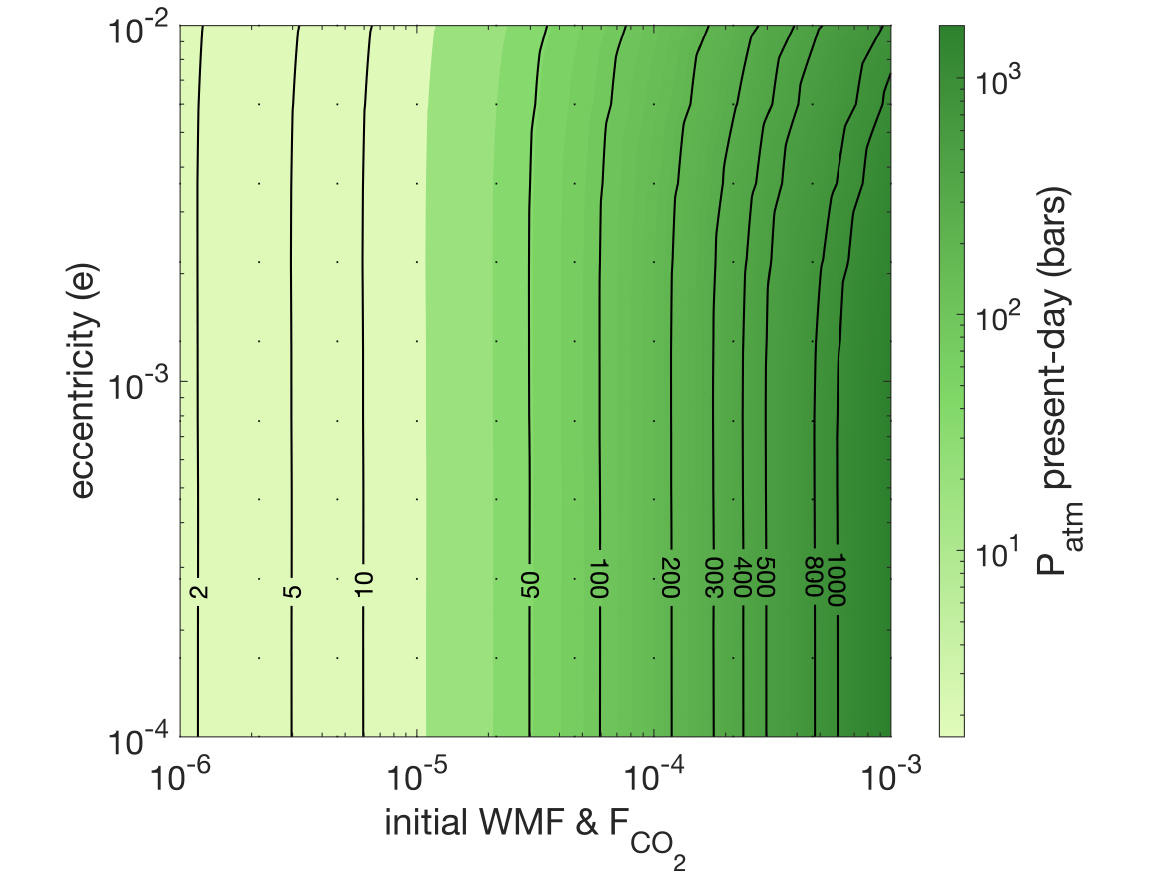}
\caption{ The predicted pressures for a tidally-enhanced present-day atmosphere on 55 Cnc e as a function of initial eccentricity and mantle volatile content. This range of initial volatile abundances was selected so that the maximum atmospheric pressure is between $\sim$2 to $\sim$1000 bars, all resulting in a $\text{CO}_2$-dominant atmosphere. Note that the black grid points denote the 100 grid simulations, and the combined initial volatile parameter represents equal parts of $F_{\text{H}_2\text{O}}$ and $F_{\text{CO}_2}$ together.}
\label{fig:presentday_tidal_Patm}
\end{figure}

We predict that an observable present-day atmosphere on 55 Cnc e must have started with an initial volatile reservoir $< 0.1$ wt\% for both $F_{\text{H}_2\text{O}}$ and $F_{\text{CO}_2}$ in Figure \ref{fig:presentday_tidal_Patm}, netting a combined total of $0.2$ wt\% for the planetary volatile mass fraction, given that atmospheric retrievals estimate a total pressure of less than $\sim$1000 bars \citep{hu_secondary_2024}. However, this prediction does not consider non-hydrodynamic mass loss mechanisms, such as tidal erosion or impact erosion, or the dissociation of $\text{CO}_2$. We discuss the possible effects of secondary mass loss mechanisms beyond hydrodynamic drag in Section \ref{sec:discussion_mass_loss}.

Note that the eccentricity tracks we employ are held constant for simplicity, due to the complex relationships between eccentricity pumping mechanisms in multi-planetary systems and tidal damping systems \citep{bolmont_emeline_tidal_2013, puranam_chaotic_2018, becker_coupled_2020, papaloizou_orbital_2021, greklek-mckeon_tidally_2025} and the ongoing uncertainties of modeling circularization timescales for the USPs and close-in planets such as 55 Cnc e \citep{mello_tidal_2025}. We discuss the ramifications of circularization timescales and eccentricity excitation on tidally-enhanced outgassing in Section \ref{sec:discussion_tidal_model}, however we expect longer circularization timescales, which will result in more prolonged heating, will also result in prolonging outgassing, which may be several magnitudes higher in the midlife evolution of the planet, at times when heating rates decline and more solidification (and therefore outgassing) occur.

\section{Discussion}\label{sec:discussion}

\subsection{XUV Flux Dependencies and CO$_2$ Photochemistry}\label{sec:discussion_photochemistry}

In our model, atmospheric mass loss is directly coupled to the strength of XUV flux evolution on the planet. The XUV flux on 55 Cnc e has been modeled from flux observations \citep{sanz-forcada_j_estimation_2011}, but the stellar model in this study uses a higher XUV flux. We also adopt an energy-limited escape model to drive a faster loss of the atmosphere. However, it has been proposed that a higher XUV flux, alongside ionized winds, can instead produce photodissociation of CO$_2$ into CO and O, which may then dissociate further into atomic components and influence the effective absorption of XUV energy. We do not consider non-thermal forms of escape on mass loss and dissociation for $\text{H}_2\text{O}$ or $\text{CO}_2$ in our model.
 
If XUV-driven CO$_2$ photolysis is effective on 55 Cnc e, loss of a CO$_2$-rich atmosphere would be enhanced, possibly resulting in an upper atmosphere saturated in CO, H, and O \citep{hu_secondary_2024}. However, delayed outgassing due to tidal heating could preserve mantle volatiles for later atmospheric formation even if mass loss greatly reduces an early atmosphere. Additionally, \citet{tian_thermal_2009} find that more than 1000 times the XUV flux of Earth may still not be sufficient to drive CO$_2$ loss on massive super-Earths of 6-10 $M_\oplus$. However, the extreme insolation flux on 55 Cnc e and other USPs far exceeds surface fluxes on habitable zone terrestrial planets, warranting further study of CO$_2$ loss and dissociation on close-in planets.

Additionally, XUV flux would be variable due to changing orbital distance for eccentric orbits that exhibit tidal heating \citep{milankovic_canon_1969, dobrovolskis_insolation_2013}. However, eccentricities would need to be large ($e>0.4$) for XUV-flux to have a discernible effect, at least for habitable zone planets \citep{palubski_habitability_2020}. For 55 Cnc e ($e < 0.01$), we assume that eccentricity-dependent variations in XUV fluxes are negligible.

\subsection{Secondary Atmospheric Mass Loss Mechanisms}\label{sec:discussion_mass_loss}

We focus on XUV-driven hydrodynamic drag rather than Jeans escape as the main thermal loss mechanism due to the extreme surface temperatures on 55 Cnc e, which is expected to drive a hydrodynamic escape regime for low values of the Jean's parameter, $\lambda_{\text{esc}} \propto {\mu_i}/{T}$. Quantifying the transition between the two thermal loss regimes has been an important consideration for some recent models \citep{volkov_thermally_2011, gronoff_atmospheric_2020, evans_thermally_2025}. In the case of extreme dissociation, the molecular weight of the escaping species is lowered, which reduces the Jean's parameter, so that hydrodynamic drag rate may become even more favored.

We model a well-mixed atmosphere containing CO$_2$, H$_2$O, and O$_2$, where hydrogen produced by water photolysis is assumed to power hydrodynamic drag of the heavier atomic and molecular species. Recent studies have shown that the interactions between multiple different molecular species via diffusion and radiative cooling can reduce the overall rate of hydrodynamic escape \citep{yoshida_sluggish_2020, yoshida_hydrodynamic_2021,yoshida2022less}, but the complex dependence on gas composition is still not well studied. Though multi-component hydrodynamic models have been used to study the rate of hydrodynamic escape from early Earth, Mars, and some exoplanets \citep[e.g.,][]{tian_hydrodynamic_2008, tian_thermal_2009}, additional work is needed to determine the extent of diffusion-limited escape through a heavier multi-species atmosphere in conjunction with a photochemical upper atmosphere model, which would be relevant to the environments found on the USPs and close-in planets. It is therefore plausible that our energy-limited escape model overestimates the loss of hydrogen and production of O$_2$ while underestimating CO$_2$ loss. The extent of photodissociation and diffusion-limited escape should be assessed in future studies. 

Although tidal forces may also enhance atmospheric mass loss rates \citep{erkaev_n_v_roche_2007}, we do not consider the effects of tidal erosion or perturbation in causing Roche lobe overflow. \citet{koskinen_mass_2022} find that a radius of 3 $R_\oplus$ is the approximate transition zone for tidal mass loss to exceed XUV-driven thermal escape on sub-Jovian planets at near-USP to USP planet orbits, which is slightly larger than the measured radius of 55 Cnc e. Quantifying the extent of Roche lobe overflow for planets around the size of 55 Cnc e, including sub-Neptunes and super-Earths, is necessary to establish the extent of atmospheric evolution and retention on the USPs and magma ocean planets in the future.

\subsection{Silicate Vapor in the Deep Atmosphere}

Silicate vapor atmospheres have been theorized to exist on some close-in rocky planets where insolation-driven surface temperatures are sufficient to drive evaporation of a magma ocean \citep{schaefer_chemistry_2009, schaefer_vaporization_2012, kite_atmosphere-interior_2016}. The high surface temperatures found on 55 Cnc e may be amenable to a lower silicate atmosphere hidden beneath a CO-$\text{CO}_2$ atmosphere \citep{hu_secondary_2024, zilinskas_characterising_2025}.  We assume a well-mixed atmosphere for our mass loss model, though a heavier lower atmosphere formed by silicate vapors could inhibit escape by hampering diffusion of light species to the upper atmosphere \citep{misener_importance_2022, misener_atmospheres_2023}. Some retrievals from JWST/NIRCam support hints of a silicate vapor atmosphere \citep{patel_jwst_2024}, however recent analysis from \citet{hu_secondary_2024} and \citet{zilinskas_characterising_2025} using JWST's MIRI and NIRCam instruments do not support a conclusion for a vaporized silicate atmosphere at the observable altitudes. However, silicate cloud formation and feedback with the surface magma ocean has been a suggested mechanism, alongside stochastic outgassing, to explain the phase-curve variability on 55 Cnc e \citep{meier_valdes_investigating_2023, loftus_extreme_2025}. Observations are not yet sufficient to rule these mechanisms out. For our model results, silicate vapor production may reduce the total outgassing of volatile species by increasing the pressure at the atmosphere-magma ocean interface \citep{fegley2016solubility}, which could effect our predictions of total atmospheric pressure for a given initial volatile content, but would help to further extend outgassing timescales by delaying volatile loss. 

\subsection{Implications of a Primordial H/He Atmosphere}

It is likely that the extreme XUV received by a young 55 Cnc e would drive the loss of a primordial H/He envelope. Although we do not model an early H/He envelope, it is possible that $\text{H}_2$ could have prolonged the magma ocean lifetime by reducing the early outgassing and escape of CO$_2$ and H$_2$O during the star's early XUV-active period \citep{schaefer_predictions_2016}. If 55 Cnc e initially formed as a hot Jupiter and migrated to its current orbital distance, both tidally-driven Roche lobe overflow and XUV-driven hydrodynamic drag may have contributed to quick stripping of the primordial atmosphere \citep{koskinen_mass_2022}. Loss of the initial H/He envelope may have then resulted in elevated rates of hydrodynamic drag, driving enhanced loss of the early outgassed atmosphere. Because of this possibility, we calculated not just the early outgassing rates, but also outgassing rates at much later times in Figure \ref{fig:Tidal_overview} and \ref{fig:Combined_Tidal_Grid}, which should be mostly independent from initial loss of the outgassed species. However, \citet{krissansen-totton_erosion_2024} found high amounts of $\text{H}_2$ could reduce mantle iron species, which would limit the final loss of oxygen. \citet{charnoz_effect_2023} also estimated <$10^{-6}$ H mass fraction could produce large amounts of $\text{H}_2\text{O}$ vapor while promoting the evaporation of a silicate atmosphere instead. Though escaping hydrogen and a silicate-dominant atmosphere on 55 Cnc e have already been mostly ruled out for the present-day \citep{zhang_no_2021, hu_secondary_2024, zilinskas_characterising_2025}, a primordial H/He envelope could have significantly altered the volatile outgassing and atmospheric chemistry of the early magma ocean and atmosphere.

\subsection{Global Versus Hemispherical Magma Ocean on 55 Cnc e}

Observations by Spitzer/IRAC have shown extreme surface temperature variation between the dayside and nightside of 55 Cnc e due to tidal-locking and intense stellar insolation \citep{demory_map_2016, mercier_revisiting_2022}, which has sometimes been interpreted as a hemispherical day-side magma ocean. Our model adopts a global magma ocean model due to the possibility of a thick greenhouse atmosphere that efficiently redistributes heat between the two hemispheres \citep{hu_secondary_2024}. Even in the case of a hemispherical magma ocean with a thin to no-atmosphere, it may be possible for the dayside magma ocean to extend to the CMB, depending on the melt composition of the mantle \citep{boukare_deep_2022}. In addition, 2D models of 55 Cnc e's interior have shown thermal convection from the relatively cool nightside can form large mantle plume upwellings on the dayside \citep{meier_tobias_g_interior_2023}. Tidal heating rates can also be affected by uncertainties relating to the response of the Love number to frictional dissipation due to an uneven radius of solidification \citep{farhat_tides_2025}. It is possible that differences in hemispherical outgassing, as a consequence of the formation of a dayside hemispherical magma ocean, with a shoreline near the terminator \citep{boukare_deep_2022}, contribute to the observed phase-curve variability. This effect would plausibly be strengthened by tidally-enhanced outgassing such as we have explored her. 

\subsection{Eccentricity Tracks and Tidal Model Limitations}\label{sec:discussion_tidal_model}

For simplicity, we do not attempt to model the tidal evolution of 55 Cnc e and instead assume constant eccentricity tracks (see Fig. \ref{fig:Tidal_overview}) and therefore constant tidal heating rates. However, \citet{bolmont_emeline_tidal_2013} and \citet{mello_tidal_2025} have demonstrated that the circularization timescales for 55 Cnc e are so short ($< 3$ Myr) that they favor near-zero eccentricity. \citet{mello_tidal_2025} suggest that the large observed eccentricity of 55 Cnc e could be generated by an undetected planet in a close resonance with planet e, which could help maintain a high tidal heating rate on long timescales, or alternatively, due to a recent chance encounter with another object that has recently pumped the eccentricity to the high observed value. In the latter case, circularization would proceed rapidly, and the planet would eventually fall onto the star in 10-100 Myr and would have experienced high tidal heating rates only in the recent time period following the encounter. In the former case, our constant tidal heating model may be a close representation of the planet's history, but in the latter, our models would predict significantly more tidal heating over the planet's lifetime than has occurred. In that case, our calculated outgassing rates for a given eccentricity track could be interpreted as the maximum present-day atmosphere formation rate possible for a system that converges to that very same eccentricity.

Additionally, our model assumes a fixed orbital and rotational state for 55 Cnc e, leaving spin-orbit coupling beyond the scope of this work, though its early dynamical history may have included inward migration and resonance-driven excitation that would alter spin-orbit evolution and contribute additional heating from asynchronous tides. However, migration and orbit evolution models imply spin-orbit synchronization occurs on $<0.1$ Myr timescales \citep{murray_solar_2000, hansen_potentially_2015}. In any case, early asynchronous tidal heating would extend the initial outgassing episode, while mature outgassing from long-term tidal heating occurs at much later timescales when asynchronous tides are negligible.

Though we assume a solid-state tidal heating model (Eq. \ref{eqn:tidal_heating}), \citet{farhat_tides_2025} applied a coupled solid-fluid treatment of tides to demonstrate that tidal forces on close-in magma ocean worlds would preclude spin-orbit resonances while shortening circularization timescales even further. Using a solid-state tidal heating model, our calculated surface temperatures still match the predicted range given by the fluid tidal model \citep{farhat_tides_2025}. However, we note that the model of \citet{farhat_tides_2025} assumes a shallow global magma ocean, rather than a very deep magma ocean, and neglects the effect of crystal formation on the dissipation rate during cooling events. Further work is needed to fully explore tidal heating in thermally evolving magma oceans.

\subsection{Oxygen Fugacity and a Reduced Mantle}\label{sec:discussion_reduced_mantle}

 Some studies have proposed CO to be the dominant background gas on 55 Cnc e \citep{hu_secondary_2024}. CO is likely to be more dominant than CO$_2$ in reduced environments, and may outgas more commonly than CO$_2$ for reduced oxygen fugacities $f_{\text{O}_2}$ \citep{sossi_redox_2020, gaillard_redox_2022}, with outgassing speciation shown to shift from $\text{CO}_2$ to $\text{CO}$ for fO$_2$ $\sim\Delta\text{IW}$$-3$ from a bulk silicate Earth (BSE) composition at 1 bar and 1400 K \citep{lichtenberg_geophysical_2022}. As discussed by \citet{hu_secondary_2024}, the lifetime of an outgassed CO atmosphere with minor CO$_2$ (< 10 bars) would be on timescales of Myrs because CO is more susceptible to hydrodynamic drag than CO$_2$ \citep{modirrousta-galian_diffusion_2024}.  Outgassing of water vapor (Fig. \ref{fig:Tidal_overview} and \ref{fig:Combined_Tidal_Grid}), could also supply the necessary oxygen to oxidize CO, leading to the production of a $\text{CO}_2$ atmosphere. Recently, reanalysis of JWST's NIRCam and MIRI data points to $\text{CO}_2$ being the favored atmospheric species in place of CO \citep{zilinskas_characterising_2025}. Regardless, without prolonged outgassing, an atmosphere is likely less observable for both CO-dominant or dissociated $\text{CO}_2$ atmospheres.


\section{Conclusions and Implications}\label{sec:conclusion}

Using our model for 55 Cnc e, we find significant implications for sustained outgassing of a secondary atmosphere in older planetary systems. Our results can be broadly summarized in the following findings:

(i). Without tidal heating, the initial planetary volatile inventory regulates the duration of outgassing and cooling of the magma ocean due to the greenhouse effect. An initial planetary mass fraction of $F_{\text{H}_2\text{O}}=0.05$ (5 wt\%) or $F_{\text{CO}_2}=0.03$ (3 wt\%) can prolong outgassing by $\sim$10 Myr. At most, the combined reservoir of both volatiles, $F_{\text{CO}_2}=0.10$ ($10\%$) and $F_{\text{H}_2\text{O}}=0.10$ (10 wt\%), can extend cooling and continuous outgassing for 30 Myr.

(ii.) With tidal heating, a trade-off between enhanced outgassing and maximum atmospheric pressures emerges. Tidal heating can maintain a high planetary thermal equilibrium, which slows cooling rates and therefore prolongs outgassing. However, low eccentricity (low tidal heating) regimes allow for more complete cooling, which results in larger overall outgassing over the planet's lifetime and higher final atmospheric pressures.

(iii.) Outgassing is broadly classified in three stages during the planet's lifetime. (1.) $<1$ Myr, where most of the early outgassed atmosphere is formed as the planet crystallizes rapidly from initial accretion. (2.) $\sim$1 Myr to $\sim$5 Gyr, stronger outgassing occurs, which increases with eccentricity at lower initial volatile abundances. However, at higher initial volatile inventories, the greatest outgassing is focused around $e\sim0.001$ due to varying thermal evolution rates. (3.) $>5$ Gyr, where outgassing rates invert to favor lower eccentricities (lower tidal heating) with volatile-rich interiors. This is because a lower present-day equilibrium temperature allows for additional outgassing during crystallization, provided that the volatile reservoir is not yet desiccated. Generalizing to the other USPs, transitions from one regime to the next would be case-dependent but expected to follow the same trend. 

(iv.) We predict the present-day atmosphere and surface of 55 Cnc e to be governed by a globally thick $\text{CO}$-$\text{CO}_2$ atmosphere, most likely $<1000$ bars, which supports a $\sim$3200 K global surface magma ocean with a nominal eccentricity of only $e=0.005$. A $\sim$2500 K magma ocean can be maintained with $0.0001<e<0.005$. A $\text{CO}_2$-dominated atmosphere with a surface pressure below 100 bars can be created with an initial $F_{\text{CO}_2}$ less than $10^{-4}$. A steam or oxygen atmosphere is disfavored due to the high XUV radiation driving hydrodynamic loss.

(v.) Observationally, we find two significant outcomes can arise from our findings. First, continued minor outgassing, driven by tidal heating, could be partly responsible for the observed phase-curve variability. Our results demonstrate that residual cooling of the primordial magma ocean, sustained by tidal heating, fuels ongoing outgassing in the present-day. Future work should investigate effects of mantle heterogeneity, which could potentially cause episodic mantle upwelling \citep{koppers_mantle_2021, gulcher_narrow_2022, meier_tobias_g_interior_2023, weis_earths_2023}, which could plausibly explain secondary eclipse variability linked to local outgassing. Moreover, efforts to detect an outgassed atmosphere on rocky USPs should consider the shift from tidal-heating-dominated to volatile-inventory-dependent outgassing regimes, which correlate approximately with planetary age.

55 Cnc e stands as a prime candidate to investigate the convergence of tidal heating, magma ocean dynamics, secondary atmosphere creation, and thermal mass loss processes, offering a coherent framework for understanding their coupled natures. Future modeling work should prioritize interactions between tidal heating and interior evolution, which may have significant implications for the fates of secondary atmospheres and surface thermal evolution.

\bibliography{55_Cnc_e}

\end{document}